\documentclass[a4,preprint,review]{elsarticle}    

\usepackage{graphicx} 
\usepackage{amsmath} 
\usepackage{amssymb}  
\usepackage{amsfonts}
\usepackage{color}
\usepackage[latin1]{inputenc}
\usepackage{psfrag}
\usepackage{eurosym}
\usepackage{booktabs}
\usepackage{algorithm}
\usepackage{algorithmicx}
\usepackage{algpseudocode}
\usepackage{multirow}
\usepackage{pdflscape}
\usepackage{enumitem}
\biboptions{numbers,sort&compress} 

\usepackage{a4wide}  

\newtheorem{problem}{Problem}
\newtheorem{remark}{Remark}

\newtheorem{definition}{Definition}

\newcommand{\ba}{\begin{array}}
\newcommand{\ea}{\end{array}}

\newcommand{\ds}{\displaystyle}

\newcommand{\Tt}{\underline{T}}
\newcommand{\TT}{\overline{T}}

\newcommand{\bu}{\mathbf{u}}
\newcommand{\bh}{\mathbf{h}}
\newcommand{\bd}{\mathbf{e}}

\newcommand{\bU}{\mathbf{U}}

\newcommand{\bT}{\mathbf{T}}

\newcommand{\bphi}{\mathbf{\Phi}}

\newcommand{\RR}{\mathcal{R}}
\newcommand{\JJ}{\mathcal{J}}

\def\eps{{\varepsilon}}

\usepackage{color}  

\newcommand{\T}{\mathcal{K}}
\newcommand{\B}{\mathcal{B}}
\newcommand{\I}{\mathcal{I}}
\newcommand{\R}{\mathbb{R}}

\newcommand{\bv}{\mathbf{v}}
\newcommand{\be}{\mathbf{e}}

\graphicspath{{./pics/}}
\journal{Applied Energy}

\begin{document}
\begin{frontmatter}
\title{An integrated model predictive control approach for optimal \\ HVAC and energy storage operation in large-scale buildings}
\author{Gianni Bianchini}\ead{giannibi@diism.unisi.it}    
\author{Marco Casini}\ead{casini@diism.unisi.it}              
\author{Daniele Pepe}\ead{pepe@diism.unisi.it}
\author{Antonio Vicino}\ead{vicino@diism.unisi.it}  
\author{Giovanni~Gino~Zanvettor}\ead{zanvettor@diism.unisi.it}  
\address {Dipartimento di Ingegneria dell'Informazione e Scienze Matematiche\\Universit\`a di Siena, Via Roma 56, 53100 Siena, Italy}  
\begin{abstract}
This paper deals with the problem of cost-optimal operation of smart buildings that integrate a centralized HVAC system, photovoltaic generation and both thermal and electrical storage devices. Building participation in a Demand-Response program is also considered. The proposed solution is based on a specialized Model Predictive Control strategy to optimally manage the HVAC system and the storage devices under thermal comfort and technological constraints. The related optimization problems turn out to be computationally appealing, even for large-scale problem instances. Performance evaluation, also in the presence of uncertainties and disturbances, is carried out using a realistic simulation framework.
\end{abstract}
\end{frontmatter}

\section{Introduction} \label{sec:introduction}
It is well-known that buildings represent one of the major electricity consumers worldwide. About a half of the total energy consumption is devoted to heating, ventilation and air conditioning (HVAC). While HVAC devices in older buildings are operated through simple rules, newer buildings exploit the opportunities offered by information and communication technologies to efficiently operate such appliances \cite{AFRAM2014APT}. An efficient control of most of the building devices yields to a global reduction of energy consumption, with a consequent reduction of emissions and energy bills \cite{LEE2016}.

Although a large part of the currently deployed building management systems are rule-based, Model Predictive Control (MPC) is gaining a lot of importance, owing to its flexibility and its ability to take  a number of different requirements and constraints into account \cite{AFRAM2014BE}.
 Indeed, to optimize the building operation cost, several applications of MPC can be found in the literature, where both linear \cite{cole_acc13} and non-linear \cite{borrelli_tcst12} dynamics are considered. In \cite{Kwak15}, an MPC-based enthalpy control algorithm has been derived. Other works deal with uncertainty in models and/or exogenous variables by using stochastic MPC approaches. Specifically, in \cite{oldewurtel12,Oldewurtel14} a tractable reformulation using chance constraints has been devised, while a scenario-based approach has been exploited in \cite{Ma15}.

In order to reduce the computational complexity which often affects MPC implementations, in \cite{DRGONA18} a machine learning approach based on multivariate regression and dimensionality reduction algorithms has been proposed, and a case study involving a 6-zone building has been carried out. In \cite{KILLIAN18}, a cooperative Fuzzy MPC has been  implemented on a real building composed of five zones.

Due to the relevance of buildings as players in smart grids, an important aspect recently considered in the building management literature  is participation in  Demand-Response (DR) programs, which enables end-users to become active players in the electricity system \cite{DR_review_pesgm07,DR_review_pesgm11,DR_review_isgt11}. DR is usually managed by an intermediary player (known as the aggregator \cite{BCVE11}) whose role is to gather flexibility from its affiliated consumers, in order to build services to be sold either to the grid operator or on the wholesale market. In response to the aggregator's  requests, which become known in advance, consumers may choose to adjust their consumption patterns in order to comply  and receive a monetary reward \cite{Chassin15, Patteeuw15}. Reduction of energy cost in presence of DR has been studied in \cite{BEHBOODI18} via an agent-based modeling approach.

Considering the growing importance of distributed generation and the relevant share of energy consumption by buildings, energy storage systems are expected to become increasingly used. Such systems are employed to achieve high levels of self-sufficiency, as well as system resiliency \cite{Loevenbruck16,Williams12}.  Furthermore, optimal sizing and efficient management of electrical storage in smart buildings, mainly in the presence of photovoltaic (PV) generation \cite{WU2017En} and plug-in electric vehicles \cite{WU2017Po}, is a widely investigated topic. In view of the participation of a building in DR programs, the presence of energy storage facilities contributes to achieve  a high level of consumption flexibility.

\subsection{Paper contribution}
The aim of this work is to propose a novel MPC-based control strategy aimed at the minimization of the electricity bill in a large-scale building.
The building is assumed to be equipped with a central HVAC system powered by heat pumps (HPs), a PV plant, thermal (TES) and electrical (EES) storage devices, and to participate in a DR program. In the proposed approach, heat pumps, heater actuators, storage devices, and PV generation facilities are managed in an optimal fashion by a centralized control unit.
 The employed DR paradigm consists of price-volume signals sent by an aggregator, which specify a monetary reward granted to the building operator upon the fulfillment of energy consumption constraints within a given time period  \cite{pekka_energycon12,APEN1}.
 
 The control law for the different actuating devices of the overall HVAC system is computed through a receding horizon algorithm involving a two-step optimization strategy. Several constraints are taken into account, namely thermal comfort, DR programs, operating limits of heat exchangers, PV power availability, storage capacity and maximum charging/discharging rates. Meteorological variables, building occupancy, light and internal appliance loads are considered as exogenous disturbances, of which the availability of measured/predicted values is assumed.
 
 The main contributions of the paper in the above context are as follows.
\begin{itemize}
\item Since the computational complexity of the control algorithm is a crucial aspect when dealing with large-scale buildings, a framework is adopted in which heat flows in the different building zones are regulated by controlling the air mass flow through each fan coil. This choice, along with approximate linear modeling of the building zones, allows for formulating the optimal control problem at each step as a Linear Program (LP). Hence, the involved computational burden scales nicely  with the problem dimension and the solver provides a fast solution even for a large number of optimization variables and constraints \cite{bv04}. Actually, the presence of DR programs calls for  the introduction of binary decision variables, changing the nature of the problem from LP to Mixed Integer Linear Program (MILP). Nevertheless, the number of such variables amounts at most to few units, i.e., the number of time intervals that contain a DR request falling inside the optimization horizon and, more importantly, it is independent of the building size (i.e., the number of zones). As a result,  the computational burden of the overall optimization algorithm is affordable even for large-scale instances, making the proposed approach suitable for real-world applications with hundreds of zones. 

\item As it is well-known in the literature, multiple sources of uncertainty come into the picture when dealing with building heating/cooling control, such as modeling inaccuracies, disturbances, weather forecasting errors and possibly undersized HVAC system. For this reason, when implementing a constrained MPC strategy, the presence of such uncertainties often causes infeasibility of the optimization problem. A standard way to deal with this issue is to replace hard bounds with soft bounds on the constraints and introduce penalties in the cost function accounting for constraint violations. Unfortunately, the choice of appropriate weights for the different penalty terms becomes a formidable task, especially in view of the high number of variables/constraints involved
and the hard-to-predict ``size'' of constraint violations. To overcome these issues, this paper proposes a different approach based on a two-step optimization strategy. At each time step, first an ancillary LP is solved in order to reset the comfort constraints, by minimizing the $l_1$ norm of the bound violations. Then, the feasible overall optimization problem is solved using the modified constraints. A further advantage of this approach is that it allows for quick recovery from realistic situations in which the original comfort constraints cannot be satisfied. 
\end{itemize}
 To validate the proposed control algorithm, numerical simulations have been performed on EnergyPlus \cite{energyplus_13}, an industry-standard realistic building model simulator. In particular, performance analysis in the presence of uncertain forecasts of exogenous variables has been carried out.

The paper is organized as follows: in Section \ref{sec:overview} an overview of the proposed control architecture is presented, in Section \ref{sec:modeling} the relevant models are introduced. In Section \ref{sec:control} the control problem is formulated, its solution derived and implementation issues discussed. The test cases are presented in Section \ref{sec:test_cases} and discussed in Section \ref{sec:discussion}. Finally, conclusions are drawn in Section \ref{sec:conclusions}.

\begin{center}
	\begin{tabular}{ll} 
		\multicolumn{2}{c}{\textbf{Nomenclature}}\\
		\hline
		\textbf{Name} & \textbf{Description} \\ \hline
		\multicolumn{2}{l}{\textbf{\emph{Acronyms}}}\\
		BCVTB & Building Controls Virtual Test Bed\\
		DR & Demand-Response\\
		EES & Electrical Energy Storage\\
		FIT & Best FIT index\\
		HP & Heat Pump\\
		HVAC & Heating, Ventilation and Air Conditioning\\
		LP & Linear Program\\
		MILP & Mixed Integer Linear Program\\
		MPC & Model Predictive Control\\
		PV & PhotoVoltaic\\
		PVUSA & PhotoVoltaics for Utility Systems Applications\\
		TES & Thermal Energy System\\
		\multicolumn{2}{l}{\textbf{\emph{Mathematical notation}}}\\
		$\R$ & Real space\\
		$\B=\{0,1\}$ & Binary set\\
		$x\in\R$ & Real scalar\\
		$\mathbf x=[x_1 x_2 \dots x_m]'$ & $m-$dimensional real vector\\
		$\mathbf X=\{x_{i,j}\}$ & Real matrix with entries $x_{i,j}$\\
		${\mathbf X}'$ & Transpose of ${\mathbf X}$\\
		$\mathcal X$ & Generic set\\
		${\mathcal X}^m$ & Cartesian product of $m$ sets identical to $\mathcal X$\\
		${\mathcal X}\backslash \{x\}$ & Set $\mathcal X$ without its element $x$\\
		$\T=\{0,1,\dots\}$ & Set of discrete time indices\\
		$k\in{\mathcal K}$ & Generic time index\\
		$\I(k,\lambda)=[k,k+\lambda)\subseteq \T$ & Generic time interval\\
		$\bU(k,\lambda)=\left[\bu(k)\,\ldots\,\bu(k+\lambda-1)\right]'$ & Matrix grouping $\bu(l)$ for $l\in\I(k,\lambda)$\\
		\multicolumn{2}{l}{\textbf{\emph{Basic notations}}}\\
		$m$ & Number of building zones\\
		${\rm Z}_i$ & $i-$th zone of the building\\
		${\rm H}_i$ & Heat exchanger device of zone $i$\\
		HP$_H$, HP$_C$ & Heat pump used for heating/cooling \\
		$\tau_s$ & Sampling time\\
		$\lambda$ & Horizon length (number of time steps inside the horizon) \\
		\hline
	\end{tabular}
	
	\begin{tabular}{ll} 
		\multicolumn{2}{c}{\textbf{Nomenclature}}\\
		\hline
		\textbf{Name} & \textbf{Description} \\ \hline
		\multicolumn{2}{l}{\textbf{\emph{Zone Thermal model}}}\\		
		$h_i$ & Heat flow conveyed by ${\rm H}_i$ into zone ${\rm Z}_i$\\
		$\bh$ & Vector of $h_i$, $i=1,\ldots,m$\\
		$T_i$ & Indoor temperature of zone ${\rm Z}_i$\\
		$\bT$ & Vector of $T_i$, $i=1,\ldots,m$\\
		$[\Tt_i,\TT_i]$ & Thermal comfort range for zone ${\rm Z}_i$\\
		$\underline{\bT}$, $\overline{\bT}$ & Vector of $\Tt_i$ and $\TT_i$, $i=1,\ldots,m$\\
		$\bd$ & Vector of measurements/predictions of exogenous inputs\\
		$\bphi^{\bT}$ & Regression matrix of zone thermal model\\
		$k_\bT,k_\bh,k_\bd$ & Model order of $\bT$, $\bh$ and $\bd$ involved in $\bphi^{\bT}$\\
		$\Theta^{\bT}$ & Parameter matrix of zone thermal model\\ 
		${\cal C}^T$ & Set of constraints of zone thermal model\\
		${\cal C}$ & Set of comfort constraints\\
		\multicolumn{2}{l}{\textbf{\emph{HVAC model}}}\\
		$T^{SND}$, $T^{RET}$ & Fluid temperature at the inlet/outlet of heat exchangers\\
		$v_i$ & Heat exchanger actuation signal (commanded air flow) for ${\rm H}_i$\\
		$\bv$ & Vector of $v_i$, $i=1,\ldots,m$\\
		$\overline v_i$ & Maximum air flow rate allowed for ${\rm H}_i$\\
		$\gamma_{i}$ & Coefficient of heating performance of heat exchanger ${\rm H}_i$\\
		${\cal C}^h_{H}$, ${\cal C}^h_{C}$ & Set of constraints of HVAC model in heating/cooling mode\\
		\multicolumn{2}{l}{\textbf{\emph{Heating mode operation}}}\\
		$T^{HP_H}_{in}$, $T^{HP_H}$ & Fluid temperature at HP$_H$ inlet/outlet\\
		$T^{HP_H}_{0}$& HP$_H$ outlet temperature reference (HP$_H$ command signal)\\
		$W^{HP_H}$& HP$_H$ electrical energy consumption in heating mode\\
		$T^{TES}$& TES fluid temperature\\
		$\alpha^{HP_H}$ & Coefficient of HP$_H$ energy consumption\\
		$\bphi^{TES}$ & Regression vector of the heating mode operation model\\
		$\Theta^{TES}$ & Parameter vector of the heating mode operation model\\ 
		$k_{TT},~k_{TH},~k_{T \bh}$ & Model order of $T^{TES}$, $T^{HP_H}$ and $\bh$ involved in $\bphi^{TES}$\\
		${\cal C}_H$ & Set of constraints of overall HVAC model in heating mode\\	
		\hline
	\end{tabular}
	
	\begin{tabular}{ll}
		\multicolumn{2}{c}{\textbf{Nomenclature}}\\
		\textbf{Name} & \textbf{Description} \\ \hline
		\multicolumn{2}{l}{\textbf{\emph{Cooling mode operation}}}\\		
		$T^{HP_C}_{in}$, $T^{HP_C}$ & Fluid temperature at HP$_C$ inlet/outlet\\
		$T^{HP_C}_{0}$& HP$_C$ outlet temperature reference (HP$_C$ command signal)\\
		$W^{HP_C}$& HP$_C$ electrical energy consumption in cooling mode\\
		$\alpha^{HP_C}$ & Coefficient of HP$_C$ energy consumption\\
		$\bphi^{C}$ & Regression vector of the cooling mode operation model\\
		$\Theta^{C}$ & Parameter vector of the cooling mode operation model\\ 
		$k_{TC},~k_{HC},~k_{C \bh}$ & Model order of $T^{HP_C}_{in}$, $T^{HP_C}$ and $\bh$ involved in $\bphi^{C}$\\
		${\cal C}_C$ & Set of constraints of overall HVAC model in cooling mode\\
		\multicolumn{2}{l}{\textbf{\emph{Electrical storage model}}}\\
		$E^{EES}$ & Battery state of charge\\
		$\overline E^{EES}$ & Battery capacity\\ 		
		$W^{EES}_+$, $W^{EES}_-$ & Battery charge/discharge signal\\
		$W^{EES}$ & Energy exchanged by the battery\\
		$\overline W^{ESS}_+$, $\overline W^{ESS}_-$ & Battery maximum charge/discharge rate\\
		$\eta$ & Battery charging/discharging efficiency\\
		${\cal C}^{EES}$ & Set of constraints of EES\\
		\multicolumn{2}{l}{\textbf{\emph{Energy consumption and DR model}}}\\
		$W$ & Energy drawn from the grid by the building\\
		$p$ & Unit energy price\\
		$W(k,\lambda)$ & Total energy drawn from the grid within $\I(k,\lambda)$\\
		$C(k,\lambda)$ & Total cost of energy within $\I(k,\lambda)$\\
		${\mathcal R}_j$ & $j-$th DR request\\
		$S_j$ & Total energy bound associated to ${\mathcal R}_j$\\
		$R_j$ & Monetary reward associated to ${\mathcal R}_j$\\
		${\mathcal P}$ & DR program (a sequence of DR requests)\\
		${\mathcal P}(k,\lambda)$ & Set of DR requests that occur within $\I(k,\lambda)$\\
		$\JJ(k,\lambda)$ & Set of indices $j$ associated to $\RR_j\in {\mathcal P}(k,\lambda)$\\
		$\epsilon_j$ & Binary variable associated to the fulfillment of ${\mathcal R}_j$\\
		$C^{\mathcal P}(k,\lambda)$ & Overall cost of operation under $\mathcal P$ within $\I(k,\lambda)$\\
		$M$ & Upper bound to the total building energy consumption per time step\\
		${\cal C}^{DR}(k,\lambda)$ &  Set of constraints of DR model within $\I(k,\lambda)$\\
		\hline
	\end{tabular}
	
	\begin{tabular}{ll} 	
		\multicolumn{2}{c}{\textbf{Nomenclature}}\\
		\textbf{Name} & \textbf{Description} \\ \hline
		\multicolumn{2}{l}{\textbf{\emph{PV generation model}}}\\
		$W^{PV}$ & Energy drawn from the PV plant\\
		$\overline W^{PV}$ & Maximum energy production of the PV plant\\
		$I$ & Global solar irradiance\\
		$T^A$ & Outside air temperature\\
		$\theta^{PV}_1,~\theta^{PV}_2,~\theta^{PV}_2$ & Parameters of PVUSA model\\
		$\Theta^{PV}$ & Parameter vector of the PV model\\ 
		${\cal C}^{PV}$ & Set of constraints of PV model\\	
		\multicolumn{2}{l}{\textbf{\emph{Optimization problem (common)}}}\\
		$h_i^*$, $v_i^*$ & Optimal value of $h_i$ and $v_i$\\
		$\hat{\bd}$ & Forecasts/measurements of exogenous inputs\\
		$\underline{\delta_i},~\overline{\delta_i}$ & Slack variables\\		
		$\Delta(k,\lambda)$ & Set including $\underline{\delta_i},~\overline{\delta_i}$ within $\I(k,\lambda)$\\
		${\mathcal C}_\Delta$ & Relaxed comfort constraint set\\
		$\Delta^{feas}(k,\lambda)$ & Optimal value of $\Delta(k,\lambda)$ guaranteeing feasibility\\
		$\star$ (subscript) & $\star=H$ denotes heating mode; $\star=C$ denotes cooling mode\\
		\multicolumn{2}{l}{\textbf{\emph{Optimization problem (heating mode)}}}\\
		$\bu_H$ & Vector grouping all command variables (heating mode operation)\\
		$\bU_H(k,\lambda)$ & Vector grouping all $\bu_H$ within $\I(k,\lambda)$\\
		$\bu_H^{opt}$, $\bU_H^{opt}(k,\lambda)$ & Optimal value of $\bu_H$ and $\bU_H(k,\lambda)$\\
		$\bT_m$ & Sensor measurement of $\bT$\\
		$T^{TES}_{m}$, $T^{HP_H}_{m}$, $E^{EES}_m$ & Sensor measurement of $T^{TES}$, $T^{HP_H}$ and $E^{EES}$\\
		\multicolumn{2}{l}{\textbf{\emph{Optimization problem (cooling mode)}}}\\
		$\bu_C$ & Vector grouping all command variables (cooling mode operation)\\
		$\bU_C(k,\lambda)$ & Vector grouping all $\bu_C(l)$ within $\I(k,\lambda)$\\
		$\bu_C^{opt}$, $\bU_C^{opt}(k,\lambda)$ & Optimal value of $\bu_C$ and $\bU_C(k,\lambda)$\\
		$T^{HP_C}_{in,m}$, $T^{HP_C}_{m}$ & Sensor measurement of $T^{HP_C}_{in}$ and $T^{HP_C}$\\
		\multicolumn{2}{l}{\textbf{\emph{Test cases}}}\\
		$\mathbf{G}'$ & Vector of internal heat gains of the building zones\\
		$d$ & Noise signal affecting exogenous variables\\ 
		$\alpha_1$, $\alpha_2$ & Model coefficients of noise signal $d$\\
		$\eps$ & Zero-mean Gaussian distributed random variable\\
		$T_{true}^A$, $\widehat T^A$ & Real/forecast outdoor temperature\\
		\hline
	\end{tabular}
\end{center}

\section{System architecture overview}\label{sec:overview}
This paper focuses on a building composed of several zones and equipped with a central HVAC system. Heating and cooling power is provided by electrical heat pumps. The heating pump HP$_H$ is assumed to be connected to a thermal storage device (TES), while the cooling pump HP$_C$ is not. Although thermal storage is quite common for both heating and cooling use, it is here employed in the first case only, with the aim of illustrating the performance of the proposed control strategy both in the presence and in the absence of a storage device.  Heat exchange at zone level is provided by fan coil units, one for each zone, exchanging heat with the TES at constant water throughput. Heat flow regulation is achieved by controlling the air mass flow through each fan coil. It is also assumed that each zone is equipped with a temperature sensor. The building also features short-term electrical energy storage (EES) facilities (e.g., via a rechargeable battery system) and PV generation. Participation of the building in a Demand-Response (DR) program is also assumed. Regulation of the HPs, TES, fan coil air flow, EES and PV is implemented via a digital centralized control unit that operates in discrete time with sampling period $\tau_s$.

\section{System modeling}\label{sec:modeling}
\begin{center}
	\begin{figure}
		\psfrag{AD}{$\cal P$}
		\psfrag{EHAT}{$\hat{\be}$}
		\psfrag{HP}{HP$_H$}
		\psfrag{HP2}{HP$_C$}
		\psfrag{TES}{TES}
		\psfrag{EES}{EES}
		\psfrag{EEES}{$E^{EES}$}
		\psfrag{WEES}{$W^{EES}$}
		\psfrag{FC}{H$_i$}
		\psfrag{ROOMI}{Z$_i$}
		\psfrag{THP}{$T^{HP_H}$}
		\psfrag{TTES}{$T^{SND}$}
		\psfrag{TTESIN}{$T^{RET}$}
		\psfrag{TT2}{$T^{TES}$}
		\psfrag{TTIN}{$T^{TES}_{in}$}
		\psfrag{THPIN}{$T^{HP_H}_{in}$}
		\psfrag{THP2}{$T^{HP_C}$}
		\psfrag{THP2IN}{$T^{HP_C}_{in}$}
		\psfrag{TES}{TES}
		\psfrag{PV}{PV}
		\psfrag{WPV}{$W_{PV}$}
		\psfrag{CONTR}{MPC}
		\psfrag{WHP}{$W^{HP_H}$}
		\psfrag{WHP2}{$W^{HP_C}$}
		\psfrag{T}{$\bT$}
		\psfrag{v}{$\bv$}
		\psfrag{E}{$\be$}
		\psfrag{TI}{$T_i$}
		\psfrag{VI}{$v_i$}
		\psfrag{HI}{$h_i$}
		\psfrag{THP0}{$T^{HP_H}_0$}
		\psfrag{THP02}{$T^{HP_C}_0$}
		\includegraphics[width=0.94\columnwidth]{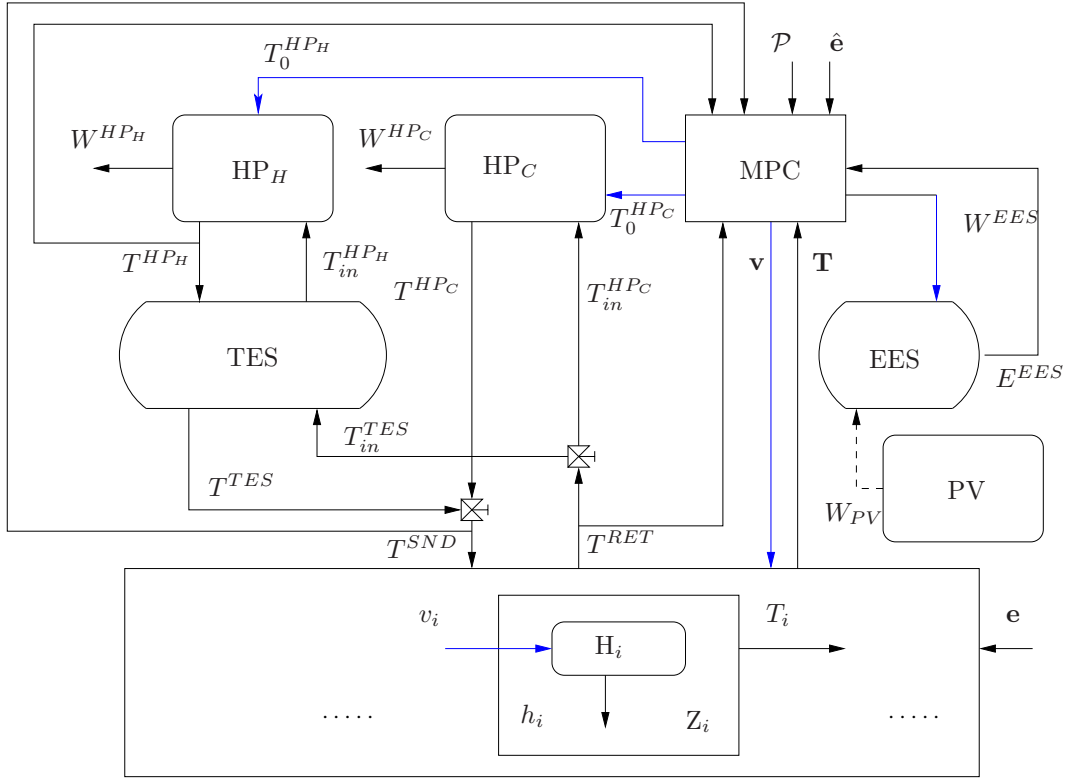}\caption{Overall HVAC and energy storage system architecture. Blue lines denote control signals. \label{fig:arch}}
	\end{figure}
\end{center}

A block diagram of the overall control system is depicted in Figure \ref{fig:arch} along with the relevant signals. The modeling of each component is detailed in the following subsections.
\subsection{Zone thermal model}\label{ss:mod}
The $m$ building zones are denoted by ${\rm Z}_1,\dots,{\rm Z}_m$, and the respective heat exchanger devices by ${\rm H}_1,\dots,{\rm H}_m$. All zones are equipped with temperature sensors, and the air flow of the heat exchangers can be independently regulated by the control unit. The control action must ensure that each room temperature satisfies a time-varying comfort constraint.\\
Let us define the following variables:
\begin{itemize}
	\item $h_i(k)\in\R$: heat flow conveyed by ${\rm H}_i$ at time $k$ into room ${\rm Z}_i$,
	\item $\bh(k)=[h_1(k) \ldots h_m(k)]'\in\R^m$,
	\item $T_i(k)$: indoor temperature of zone ${\rm Z}_i$ at time $k$,
	\item $\bT(k)=[T_1(k) \ldots T_m(k)]'\in\R^m$,
	\item $[\Tt_i(k),\TT_i(k)]$: thermal comfort range for ${\rm Z}_i$ at time $k$,
	\item $\underline{\bT}(k)=[\Tt_1(k) \ldots \Tt_m(k)]'\in\R^m$, $\overline{\bT}(k)=[\TT_1(k) \ldots \TT_m(k)]'\in\R^m$,
\end{itemize}
The dynamics of the indoor temperatures $\bT(k)$ depend on the heat flows $\bh(k)$ and on exogenous variables like
outdoor temperature, solar radiation, appliances, lighting, occupancy, etc. For the sake of simplicity,  available measurements/forecasts of some/all of such variables are collected in a vector $\bd(k)$. Assuming linear time-invariant dynamics, the zone temperature vector evolution can be modeled in regressive form as:
\begin{equation}\label{eq:Tdyn}
\bT(k+1)  =  \Theta^{\bT}~\bphi^{\bT}(k),
\end{equation}
where the regression matrix $\bphi(k)$ is given by
\begin{equation}\label{eq:Tdyn2}
\begin{array}{r}
\bphi^{\bT}(k) = [ \bT'(k)~\dots~\bT'(k-k_\bT)\\ 
\bh'(k)~\dots~\bh'(k-k_\bh) \\
\bd'(k)~\dots~\bd'(k-k_\bd)]',
\end{array}
\end{equation}
being $k_\bT,k_\bh,k_\bd$ suitable nonnegative integers that define the model order. The matrix $\Theta^{\bT}$ collects the model parameters.

According to the above, thermal comfort achievement at time $k$ is expressed by the element-wise constraint 
\begin{equation}\label{eq:comfort}
\underline{\bT}(k) \leq \bT(k) \leq \overline{\bT}(k).
\end{equation}
It is appropriate to group constraint sets \eqref{eq:Tdyn},\eqref{eq:Tdyn2},\eqref{eq:comfort} as follows:
\[
{\cal C}^T(k) = \{\mbox{\eqref{eq:Tdyn},\eqref{eq:Tdyn2}}\},~~{\cal C}(k) = \{\mbox{\eqref{eq:comfort}}\} .
\]

\subsection{HVAC system model}
The HVAC system operates at constant water flow rate. With reference to Figure \ref{fig:arch}, let us define the following quantities:
\begin{itemize}
	\item $T^{SND}(k)\in\R$: fluid temperature at the inlet of heat exchangers,
	\item $T^{RET}(k)\in\R$: temperature of return fluid,
	\item $v_i(k)\in\R$: heat exchanger actuation signal, i.e., commanded air flow at time $k$ for ${\rm H}_i$,
	\item $\bv(k)=[v_1(k) \ldots v_m(k)]'\in\R^m$,
	\item $\overline v_i$: maximum air flow rate allowed for ${\rm H}_i$, i.e.,
	\begin{equation}\label{eq:actcon}
	0\leq v_i(k)\leq \overline v_i.
	\end{equation} 
\end{itemize}
For each zone Z$_i$, the heat flow rate $h_i(k)$ can be expressed as
\begin{equation}\label{eq:hvsv}
h_i(k)=\gamma_i\left(T^{SND}(k)-T_i(k)\right) v_i(k),
\end{equation}
where $\gamma_{i}$ is a coefficient of heating performance pertaining to heat exchanger H$_i$. Note that, in view of \eqref{eq:hvsv}, enforcing limitation \eqref{eq:actcon} for all zones yields the linear constraint sets
\begin{equation}\label{eq:actcon2}
{\cal C}^h_{H}(k) = \{0\leq h_i(k) \leq \gamma_i\left(T^{SND}(k)-T_i(k)\right) \overline{v}_i,~~~i=1,\dots,m\} 
\end{equation}
and
\begin{equation}\label{eq:actcon3}
{\cal C}^h_{C}(k) = \{0\geq h_i(k) \geq \gamma_i\left(T^{SND}(k)-T_i(k)\right) \overline{v}_i,~~~i=1,\dots,m\} 
\end{equation}
when the heat exchanger operates in heating and cooling mode, respectively. 

\subsubsection{Heating mode operation with thermal storage}
The dynamics of the HVAC system in heating mode is now described. As pointed out in Section \ref{sec:overview}, the heat pump HP$_H$ is coupled to a thermal energy storage (TES). Define:
\begin{itemize}
	\item $T^{HP_H}_{in}(k)\in\R$: thermal fluid temperature at HP inlet,
	\item $T^{HP_H}(k)\in\R$: thermal fluid temperature at HP outlet,
	\item $T^{HP_H}_{0}(k)\in\R$: HP outlet temperature reference, i.e., HP command signal,
	\item $W^{HP_H}(k)\in\R$: HP electrical energy consumption within the $k-$th time step,
	\item $T^{TES}(k)\in\R$: TES fluid temperature.
\end{itemize}

The inlet temperature of each heat exchanger H$_i$ is assumed to be uniform and equal to the TES outlet temperature, i.e.,
\[
T^{SND}(k)=T^{TES}(k) .
\]
Under the realistic assumption that the time constants of the HP thermal fluid dynamics in response to a change in the reference $T^{HP}_0(k)$ are much smaller than the control sampling period $\tau_s$, the HP outlet fluid temperature can be expressed by
\begin{equation}\label{eq:HPH}
T^{HP_H}(k+1) =  T^{HP_H}_0(k),
\end{equation}
so that the HP can be modeled as a unit time step delay system.\\
A further reasonable assumption is that the heat exchange at TES level is efficient enough such that
\begin{equation}\label{eq:HPHin}
T^{HP_H}_{in}(k) = T^{TES}(k) .
\end{equation}
It is worth noting that the above assumptions are indeed satisfied by the real-world devices emulated by EnergyPlus \cite{energyplus_13}, the widely used realistic building simulation software used here for experimental validation.\\
The HP has to be switched on when the reference temperature is greater than the fluid temperature at the $HP$ inlet, i.e., as long as
$$
T^{HP_H}_0(k) > T^{HP_H}_{in}(k) = T^{TES}(k),
$$
which leads to an electrical energy consumption 
\begin{equation}\label{eq:HPpower}
W^{HP_H}(k)= \alpha^{HP_H}\left(T^{HP_H}_0(k)-T^{TES}(k)\right),
\end{equation}
where $\alpha^{HP_H}$ represents a device specification. 
Otherwise, the HP is switched off and $W^{HP_H}(k)=0$. Therefore, the condition 
\begin{equation}\label{eq:HPactivation}
T^{HP_H}_0(k) \geq T^{TES}(k)
\end{equation}
can be assumed without loss of generality.\\
Concerning the TES dynamics, it is worth observing that $T^{TES}(k)$ depends on $T^{HP_H}(k)$ and on the return water temperature $T^{RET}(k)=T^{TES}_{in}$, which in turn dynamically depends on $T^{SND}(k)=T^{TES}(k)$ and on the total heat exchange at the heater level, i.e., $h(k) = h_1(k)+\dots+h_m(k)$. Therefore, the TES temperature dynamics can be modeled in regressive form as
\begin{equation}\label{eq:Hdynamics1}
T^{TES}(k+1) = \Theta^{TES} \bphi^{TES}(k),
\end{equation}
where the regression vector $\bphi^{TES}(k)$ is given by
\begin{equation}\label{eq:Hdynamics2}\begin{array}{r}
\bphi^{TES}(k)   =  [T^{TES}(k)~\dots~T^{TES}(k-k_{TT}) \\ T^{HP_H}(k)~\dots~T^{HP_H}(k-k_{TH}) \\ \bh'(k)~\dots~\bh'(k-k_{T \bh})
]', \end{array}
\end{equation}
being $\Theta^{TES}$ the (row) parameter vector and $k_{TT},~k_{TH},~k_{T \bh}$ suitable model orders.

Given the models introduced above, and observing that the thermal energy stored in the TES is proportional to $T^{TES}(k)$, it is worth remarking that full control of individual zone temperatures and thermal energy storage can be achieved by manipulating the control variables $T^{HP_H}_0(k)$ and $\bv(k)$.

\subsubsection{Cooling mode operation}\label{subsec:cooling_mode}
In order to derive a model of HVAC system in cooling mode, let us define the following:
\begin{itemize}
	\item $T^{HP_C}_{in}(k)\in\R$: thermal fluid temperature at the inlet of the cooling heat pump,
	\item $T^{HP_C}(k)\in\R$: thermal fluid temperature at HP outlet,
	\item $T^{HP_C}_{0}(k)\in\R$: HP outlet temperature reference, i.e., HP command signal,
	\item $W^{HP_C}(k)\in\R$: HP electrical energy consumption within the $k-$th time step,
\end{itemize}

Similarly to the heating mode, the cooling HP dynamics can be modeled as: 
\begin{equation}\label{eq:HPC}
T^{HP_C}(k+1) =  T^{HP_C}_0(k) .
\end{equation}
As long as $T^{HP_C}_0(k) \leq T^{HP_C}_{in}$, the HP is switched on and the corresponding electrical energy consumption is given by
\begin{equation}\label{eq:HPpowerC}
W^{HP_C}(k)= \alpha^{HP_C}\left(T^{HP_C}_{in}(k)-T^{HP_C}_0(k)\right),
\end{equation}
where $\alpha^{HP_C}$ pertains to the given device. Otherwise, it is switched off and $W^{HP_C}(k)=0$. Therefore, it can be assumed that
\begin{equation}\label{eq:HPactivationC}
T^{HP_C}_0(k) \leq T^{HP_C}_{in}(k).
\end{equation}
In this mode, it holds that $T^{SND}(k)=T^{HP_C}(k)$ and $T^{RET}(k)=T^{HP_C}_{in}(k)$. The latter quantity dynamically depends on $T^{SND}(k)$ and on the total zone heat flow $h(k) = h_1(k)+\dots+h_m(k)$. Therefore, the following regressive model can be used for $T^{HP_C}_{in}(k)$:
\begin{equation}\label{eq:Cdynamics1}
T^{HP_C}_{in}(k+1) = \Theta^{C} \bphi^{C}(k),
\end{equation}
where the regression vector $\bphi^{C}(k)$ is given by
\begin{equation}\label{eq:Cdynamics2}\begin{array}{r}
\bphi^{C}(k)   =  [T^{HP_C}_{in}(k)~\dots~T^{HP_C}_{in}(k-k_{TC}) \\ T^{HP_C}(k)~\dots~T^{HP_C}(k-k_{HC}) \\ \bh'(k)~\dots~\bh'(k-k_{C \bh})
]' \end{array}
\end{equation}
where $\Theta^{C}$ is the (row) parameter vector and $k_{TC},~k_{HC},~k_{C \bh}$ define the model orders.

The constraint sets pertaining to the HVAC system operating in heating and cooling mode, respectively, can be grouped as follows:
\[
{\cal C}_H(k) = \{\mbox{\eqref{eq:HPH},\eqref{eq:HPpower},\eqref{eq:HPactivation},\eqref{eq:Hdynamics1},\eqref{eq:Hdynamics2}}\},~~{\cal C}_C(k) = \{\mbox{\eqref{eq:HPC},\eqref{eq:HPpowerC},\eqref{eq:HPactivationC},\eqref{eq:Cdynamics1},\eqref{eq:Cdynamics2}}\}  .
\]

\subsection{Electrical storage model}\label{ss:ESm}
The characterization of the storage system based on rechargeable batteries used in this paper is as follows (see, e.g., \cite{HITTINGER2015,AGT16}). 
Let $E^{EES}(k)$ represent the state of charge at time $k$. Denote as $W^{EES}_+(k)\geq 0$ and $W^{EES}_-(k)\geq 0$ the battery charge and discharge signals, i.e., the amount of energy fed to and drawn from the battery in the $k-$th time interval, respectively. It holds that
\begin{equation}\label{eq:EESdyn}
E^{EES}(k+1)=E^{EES}(k)+\eta W^{EES}_+(k)-\frac{1}{\eta} W^{EES}_-(k),
\end{equation}
where $0<\eta<1$ is the battery efficiency. Furthermore, the amount of battery energy exchange in the same interval reads
\begin{equation}\label{eq:EESdyn2}
W^{EES}(k) =  W^{EES}_+(k)- W^{EES}_-(k) .
\end{equation}
The following additional constraints
\begin{equation}\label{eq:EESconstr1}
0\leq  W^{EES}_+(k) \leq \overline W^{EES}_+,
\end{equation}
\begin{equation}\label{eq:EESconstr11}
0\leq  W^{EES}_-(k) \leq \overline W^{EES}_-,
\end{equation}
\begin{equation}\label{eq:EESconstr2}
0\leq E^{EES}(k) \leq \overline E^{EES},
\end{equation}
where $\overline W^{ESS}_+,\overline W^{ESS}_-$ and $\overline E^{EES}$ depend on the storage size and technology, and represent the maximum charge/discharge rate and capacity, respectively. 
The constraint sets pertaining to EES can be grouped as:
\[
{\cal C}^{EES}(k) = \{\mbox{\eqref{eq:EESdyn},\eqref{eq:EESdyn2},\eqref{eq:EESconstr1},\eqref{eq:EESconstr11},\eqref{eq:EESconstr2}} \}  .
\]
In order for the model to be consistent, $W^{EES}_+(k)$ and $W^{EES}_-(k)$ cannot both be nonzero. This can be guaranteed by introducing the nonlinear constraint $W^{EES}_+(k) W^{EES}_-(k)=0$, $\forall k$. See Remark~\ref{rem:EES} in Section~\ref{sec:control} on how the introduction of this constraint can be circumvented.

\subsection{PV generation model}\label{ss:PV}
In this work,  the well-known PVUSA model of a PV plant \cite{BIB:PVUSA} is exploited. In this model, the maximum energy that can be drawn from the plant in the $k-$th time interval is expressed as a function of irradiance and environmental temperature as follows:
\begin{equation}
\label{EQ:PVUSA}
\overline W^{PV}(k) = \theta^{PV}_1 I(k) + \theta^{PV}_2 I^2(k) +\theta^{PV}_3 I(k)T^A(k),
\end{equation}
where $I(k)$ is the global solar irradiance and $T^A(k)$ is the outside air temperature. Despite its simplicity, this model is very accurate when its parameters are fit to measured data and several efficient methods for their estimation under various conditions have been proposed (see \cite{pepe2017model},\cite{BIB:EC18},\cite{BIB:ISGT} and references therein). Without loss of generality, it can be assumed that $I(k)$,$I^2(k)$, and $I(k)T_a(k)$ are part of the exogenous variable vector $\bd(k)$. Therefore, \eqref{EQ:PVUSA} can be written in the form
\begin{equation}
\label{EQ:PVUSA2}
\overline W^{PV}(k) = \Theta^{PV} \bd(k)
\end{equation}
which represents a static linear-in-the-parameters model. 

Denoting with $W^{PV}(k)$ the amount of energy drawn from the PV plant in the $k-$th interval, the following constraint must hold:
\begin{equation}\label{eq:PV1}
0\leq W^{PV}(k)\leq \overline W^{PV}(k).
\end{equation}
The latter two equations are grouped in the constraint set
\[
{\cal C}^{PV}(k) = \{\mbox{\eqref{EQ:PVUSA2},\eqref{eq:PV1}}\} .
\]

\subsection{Energy consumption and Demand-Response model}\label{ss:DRm}
Let $W(k)$ represent the amount of energy drawn from the grid by the building in the $k-$th time interval. According to the model introduced above, the electrical energy balance equation reads:
\begin{equation}
W(k)=W^{HP}(k)+W^{EES}(k)-W^{PV}(k),
\end{equation}
with $W(k)\geq 0$, since the possibility of injecting excess energy into the grid is not currently considered. 

Let $p(k)$ represent the unit energy price or a forecast thereof, and consider a generic time horizon $\I(k,\lambda)$. The total energy drawn from the grid within $\I(k,\lambda)$ is given by
\begin{eqnarray}
W(k,\lambda)=\sum_{l=k}^{k+\lambda-1} W(l),
\end{eqnarray}
and the total expected cost of energy in the same interval is equal to
\begin{eqnarray}
C(k,\lambda)=\sum_{l=k}^{k+\lambda-1} p(l)W(l).
\end{eqnarray}

A Demand-Response model based on price-volume signals is considered in this paper. Such a model was introduced in \cite{APEN1} and is recalled next. A DR program is modeled as a sequence of DR requests ${\mathcal R}_j$, each carrying a time horizon $\I(h_j,\mu_j)$, a total energy bound $S_j$, and a monetary reward $R_j$. A request ${\mathcal R}_j$ is satisfied if the total building consumption within $\I(h_j,\mu_j)$, i.e., $W(h_j,\mu_j)$, is less or equal to the threshold $S_j$. In this case, a monetary reward $R_j$ is granted.
\begin{definition}
	A DR program ${\mathcal P}$ is a sequence of DR requests ${\mathcal R}_j$, $j=1,2,\dots$, where ${\mathcal R}_j$ is the set
	\begin{eqnarray}
	{\mathcal R}_j=\left\{\I(h_j,\mu_j),S_j,R_j\right\} ,
	\end{eqnarray}
	being $\I(h_j,\mu_j)\subseteq \T$ and $ \I(h_{j_1},\mu_{j_1})\cap \I(h_{j_2},\mu_{j_2})=\emptyset,~  \forall j_1\neq j_2$.

	The request ${\mathcal R}_j$ is satisfied if and only if
	\begin{eqnarray}
	W(h_j,\mu_j)\leq S_j .
	\end{eqnarray}
\end{definition}
For any given time horizon $\I(k,\lambda)$, let
\begin{eqnarray}
{\mathcal P}(k,\lambda)=\left\{\RR_j:\I(h_j,\mu_j)\subseteq \I(k,\lambda) \right\},
\end{eqnarray}
be the set of DR requests that occur within the time horizon. Moreover, define $\JJ(k,\lambda)$ as the set of indices identifying such DR requests, i.e.,
\begin{eqnarray}
\JJ(k,\lambda)=\left\{j:\RR_j\in {\mathcal P}(k,\lambda)\right\}.
\end{eqnarray}
To each request ${\mathcal R}_j$, let us associate a binary variable $\epsilon_j\in\B$ defined as
\begin{eqnarray}
\epsilon_j=\left\{\ba{rl}1 & {\rm if}~{\mathcal R}_j {\rm ~is~satisfied} \\ 0 & {\rm otherwise .} \ea\right.
\end{eqnarray}

The overall expected cost of operation of the building HVAC system  within the time horizon $\I(k,\lambda)$ under the DR program $\mathcal P$, is therefore given by
\begin{eqnarray}
C^{\mathcal P}(k,\lambda) = C(k,\lambda) -\sum_{j\in\JJ(k,\lambda)}\epsilon_j R_j,
\end{eqnarray}
i.e., the expected cost of energy minus the total reward for the satisfied DR requests. Let $M$ be an a-priori known upper bound to the total building energy consumption per time step. Then, the set of constraints
\begin{equation}\label{eq:DRconstr}
{\cal C}^{DR}(k,\lambda)=\left\{ W(h_j,\mu_j)\leq \epsilon_j S_j + (1-\epsilon_j) M\mu_j,~~j\in{\cal J}(k,\lambda),~~\epsilon_j\in{\cal B} \right\} 
\end{equation}
drives each  variable $\epsilon_j$ to 1 if the respective DR request can be fulfilled, and to 0 otherwise.

The reader is referred to \cite{APEN1} for further details on this model and its MPC implementation.

\section{Optimal HVAC and storage operation problem}\label{sec:control}
The goal of this section is to devise an optimal schedule for the operation of the overall system (HVAC, PV, thermal and electrical storage) in order to minimize the building electricity bill under a DR program ${\mathcal P}$, while preserving comfort constraints. On a given interval $\I(k,\lambda)$, such a problem amounts to the optimal manipulation of the control variables $\bv (l)$, $T^{HP_H}_0(l)$ $[T^{HP_C}_0(l)]$, $W^{EES}(l)$ for $l\in \I(k,\lambda)$ in order to minimize $C^{\mathcal P}(k,\lambda)$. To this purpose, it should be observed that the model is not linear in the above variables due to \eqref{eq:hvsv}. Nevertheless, the model becomes linear if $\bh(l)$ instead of $\bv(l)$ are considered as decision variables. Once the optimal values $h_i^*(l)$ of $h_i(l)$ are available for zone Z$_i$, the corresponding optimal heat exchanger actuation is given by
\begin{equation}
v_i^*(l)=\frac{h_i^*(l)}{\gamma_i\left(T^{SND}(l)-T_i(l)\right)}.
\end{equation}
In the formulation of the optimization problem, suitable forecasts or sensor measurements $\hat{\bd}(k)$ of the exogenous variables are assumed to be available. Therefore, in the sequel, $\bd(k)=\hat{\bd}(k)$.

\subsection{Heating mode operation with thermal storage}
In order to formulate the optimal operation problem in heating mode over a time interval ${\cal I}(k,\lambda)$, the set of command variables $\bU_H(k,\lambda)$, where
\[
\bu_H(l)=[\bh'(l),~T^{HP_H}_0(l),~W^{EES}(l)]',
\]
is considered. Moreover, sensor measurements $\bT_m(k)$, $T^{TES}_{m}(k)$, $T^{HP_H}_{m}(k)$, $E^{EES}_m(k)$ of $\bT(k)$, $T^{TES}(k)$, $T^{HP_H}(k)$, $E^{EES}(k)$, respectively, are assumed to be available. Hence, the problem of minimizing the total energy cost $C^{\mathcal P}(k,\lambda)$ over the time interval $\I(k,\lambda)$ can be cast as a mixed-integer linear program (MILP) as follows.
\begin{problem}\label{pb:main_problem} Optimal combined heating/storage control under DR program ${\mathcal P}$.
	\begin{eqnarray}\label{eq:main_problem}
	\ba{ll}\ds \bU_H^{opt}(k,\lambda)= \displaystyle{\arg \min_{\small \ba{c} \bU_H(k,\lambda) \\ \epsilon_j : j\in\JJ(k,\lambda) \ea} } C^{\cal P}(k,\lambda) & ~\\ 
	\mbox{\rm subjected to}\\
	{\cal C}^T(l), {\cal C}(l), {\cal C}^h_{H}(l), {\cal C}_H(l), {\cal C}^{EES}(l),  {\cal C}^{PV}(l)~~~ \forall l\in \I(k,\lambda)\\
	{\cal C}^{DR}(k,\lambda)
	\ea%
	\end{eqnarray}
\end{problem}

\subsection{Cooling mode operation}
In cooling mode, the set of control variables amounts to $\bU_C(k,\lambda)$, where
\[
\bu_C(l)=[\bh'(l),~T^{HP_C}_0(l),~W^{EES}(l)]'.
\]
 Sensor measurements $\bT_m(k)$, $T^{HP_C}_{in,m}(k)$, $T^{HP_C}_{m}(k)$, $E^{EES}_m(k)$ of $\bT(k)$, $T^{HP_C}_{in}(k)$, $T^{HP_C}(k)$, $E^{EES}(k)$, respectively, are considered.
The problem of minimizing the total energy cost over $\I(k,\lambda)$ has the following formulation.
\begin{problem}\label{pb:main_problemC} Optimal combined cooling/storage control under DR program ${\mathcal P}$.
	\begin{eqnarray}\label{eq:main_problemC}
	\ba{ll}\ds \bU_C^{opt}(k,\lambda)= \displaystyle{\arg \min_{\small \ba{c} \bU_C(k,\lambda) \\ \epsilon_j : j\in\JJ(k,\lambda) \ea} } C^{\cal P}(k,\lambda) & ~\\ 
	\mbox{\rm subjected to}\\
	{\cal C}^T(l), {\cal C}(l), {\cal C}^h_{C}(l), {\cal C}_C(l), {\cal C}^{EES}(l), {\cal C}^{PV}(l) ~~~ \forall l\in \I(k,\lambda)\\
	{\cal C}^{DR}(k,\lambda)
	~\\
	\ea%
	\end{eqnarray}
\end{problem}

\subsection{Managing constraint violations and uncertainty}\label{subsec:manage_constr}
The feasibility of the optimization problems introduced in the previous subsections cannot be guaranteed in general due to a number of circumstances that may occur in real use cases. Among such factors are the presence of uncertainties in the building component models, errors in sensor measurements or exogenous variable forecasts $\hat\bd(k)$, inappropriate sizing or faults of the HVAC system, etc. Hence, a certain degree of constraint violation must always be accepted. In order to mitigate this issue, a common approach in MPC is to resort to soft constraints, that is, removing the relevant hard constraints and replace them with penalization terms in the objective function. However, in complex scenarios such as the one dealt with in this paper, designing suitable constraint penalization coefficients to be used in the cost function can be a very difficult task, and the optimal solution of the relaxed problem may turn out to be very sensitive to such design. To overcome this issue, a two-step procedure is adopted in this work, which is based on the observation that the constraints that may actually incur violations are the comfort constraints ${\cal C}(\cdot)$. In the first step, such constraints are relaxed by introducing  a set of slack variables, and a modified version of Problem \ref{pb:main_problem} or \ref{pb:main_problemC} is solved, in which the objective to be minimized is the $1-$norm of all constraint violations over ${\cal I}(k,\lambda)$. In the second step, Problem \ref{pb:main_problem} or \ref{pb:main_problemC} is solved replacing the original comfort constraints with the optimal relaxed bounds computed in the first step. Note that this substitution always ensures feasibility of the latter problems.

In order to illustrate the procedure, the following set of positive slack variables pertaining to each time interval ${\cal I}(k,\lambda)$ is introduced:
$$
\Delta(k,\lambda) = \left\{(\underline{\delta_i}(l),\overline{\delta_i}(l)),~\underline{\delta_i}(l)\geq 0,~\overline{\delta_i}(l)\geq 0,~i=1,\dots,m,~l\in \I(k,\lambda)\right\},
$$
along with the relaxed constraint set
$$
{\mathcal C}_\Delta(k)=\left\{ \Tt_i(k)-\underline{\delta_i}(k) \leq T_i(k) \leq \TT_i(k)+\overline{\delta_i}(k) ~~\forall i=1,\dots,m \right\}.
$$
Then, the following linear program, which is a modified version of Problems \ref{pb:main_problem} and \ref{pb:main_problemC}, is considered:
\begin{problem}\label{pb:powa}
	\begin{eqnarray}\label{eq:slackpowa}
	\ba{ll} \ds \Delta^{feas}(k,\lambda)= \displaystyle{\arg \min_{\small \Delta(k,\lambda),\bU_{\star}(k,\lambda)} \sum_{l\in{\cal I}(k,\lambda),~i=1,\dots,m} \overline\delta_i(l)+\underline\delta_i(l)} & ~\\ ~\\
	\mbox{\rm subjected to}\\
	{\cal C}^T(l), {\cal C}_\Delta(l), {\cal C}^h_{\star}(l), {\cal C}_\star(l) ~~~ \forall l\in \I(k,\lambda)\\
	{\cal C}^m_{\star}(k)
	\ea%
	\end{eqnarray}
\end{problem}
where $\star=H$ for heating mode (Problem \ref{pb:main_problem}) and $\star=C$ for cooling mode (Problem \ref{pb:main_problemC}). All other constraints (battery, DR, etc.) do not influence feasibility and can be omitted. Notice that the optimum of Problem \ref{pb:powa} represents the minimum 1-norm of the constraint violations that must be accepted.

\subsection{MPC algorithm}
The proposed control procedure is implemented in a receding horizion fashion as the two-step optimization strategy in Algorithm \ref{algorithm}.
\begin{algorithm}[H]
	\caption{MPC controller implementation}\label{algorithm}
	\begin{algorithmic}[1]
		\State $k\leftarrow 0$
		\State Acquire sensor measurements  $\bT_m(k)$, $T^{TES}_{m}(k)$, $T^{HP_H}_{m}(k)$ $[T^{HP_C}_{in,m}(k)$, $T^{HP_C}_{m}(k)]$, $E^{EES}_m(k)$ and exogenous variable forecasts $\hat \bd(l)$, $l\in{\cal I}(k,\lambda)$
		\State Solve Problem \ref{pb:powa} for
		\[
		\Delta^{feas}(k,\lambda) = \left\{(\underline{\delta_i}^{feas}(l),\overline{\delta_i}^{feas}(l)),~i=1,\dots,m,~l\in \I(k,\lambda)\right\}
		\]
		\State Define the constraint set
		$$
		{\mathcal C}^{feas}_\Delta(k)=\left\{\Tt_i(k)-\underline{\delta_i}^{feas}(k)  \leq T_i(k) \leq  \TT_i(k)+\overline{\delta_i}^{feas}(k)~~\forall i=1,\dots,m \right\}.
		$$
		\State Solve Problem \ref{pb:main_problem} [Problem \ref{pb:main_problemC}] for $\bU^{opt}_\star(k,\lambda)$ setting
		\[
		{\cal C}(l) = {\mathcal C}^{feas}_\Delta(l) 
		\]
		\State Actuate $\bu_\star^{opt}(k)$
		\State $k\leftarrow k+1$ and repeat from 2:
	\end{algorithmic}
\end{algorithm}

\begin{remark}
	At each step $k$, the proposed procedure consists of the solution of an LP (Problem 3) and of a MILP (Problem 1 or 2) involving a number of binary variables equal to the number of DR requests falling inside the prediction horizon ${\cal I}(k,\lambda)$. In realistic scenarios, the number of integer variables in the MILP is at most a few units \cite{BCVE11} . The number of continuous variables as well as of constraints in all optimization problems scales in a linear fashion with the number of building zones. This makes the proposed procedure readily applicable to real-world scenarios involving large-scale buildings consisting of hundreds of zones, as demonstrated by the simulation experiments reported in Section \ref{sec:test_cases}.
\end{remark}
\begin{remark}\label{rem:EES}
	The formulation of the optimization problems involved in Algorithm \ref{algorithm} does not take into account the constraint that the electrical storage cannot charge and discharge simultaneously, i.e., that $W^{EES}_+(l)$ and $W^{EES}_-(l)$ cannot be both nonzero. However, unless $E^{EES}(l+1)=0$ or  $E^{EES}(l+1)=\overline{E}^{EES}$, such condition is always met because the battery efficiency satisfies $\eta<1$. On the contrary, when the battery state of charge hits one of the bounds, $W^{EES}(l)$ can be set such that  $E^{EES}(l+1)=0$ or $E^{EES}(l+1)=\overline{E}^{EES}$ according to \eqref{eq:EESdyn},\eqref{eq:EESdyn2}.
\end{remark}

\section{Test cases} \label{sec:test_cases}

In this section, the application of the proposed control technique to a large-scale building is presented. Heating and cooling operations are considered in Sections~\ref{subsec:winter} and \ref{subsec:summer}, respectively, while Section~\ref{subsec:uncertainty} is devoted to analyzing the performance of the proposed method when inaccurate forecasts of exogenous variable are available. A discussion of the results is given in Section~\ref{sec:discussion}.

\begin{figure}[H]
	\centering %
	\includegraphics[width=0.35\columnwidth]{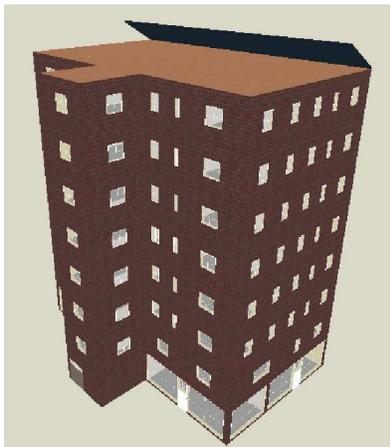}
	\caption{Rendering of the building used in test cases.}\label{fig:building}
\end{figure}

\begin{figure}[H]
	\centering %
	\includegraphics[width=0.3\columnwidth]{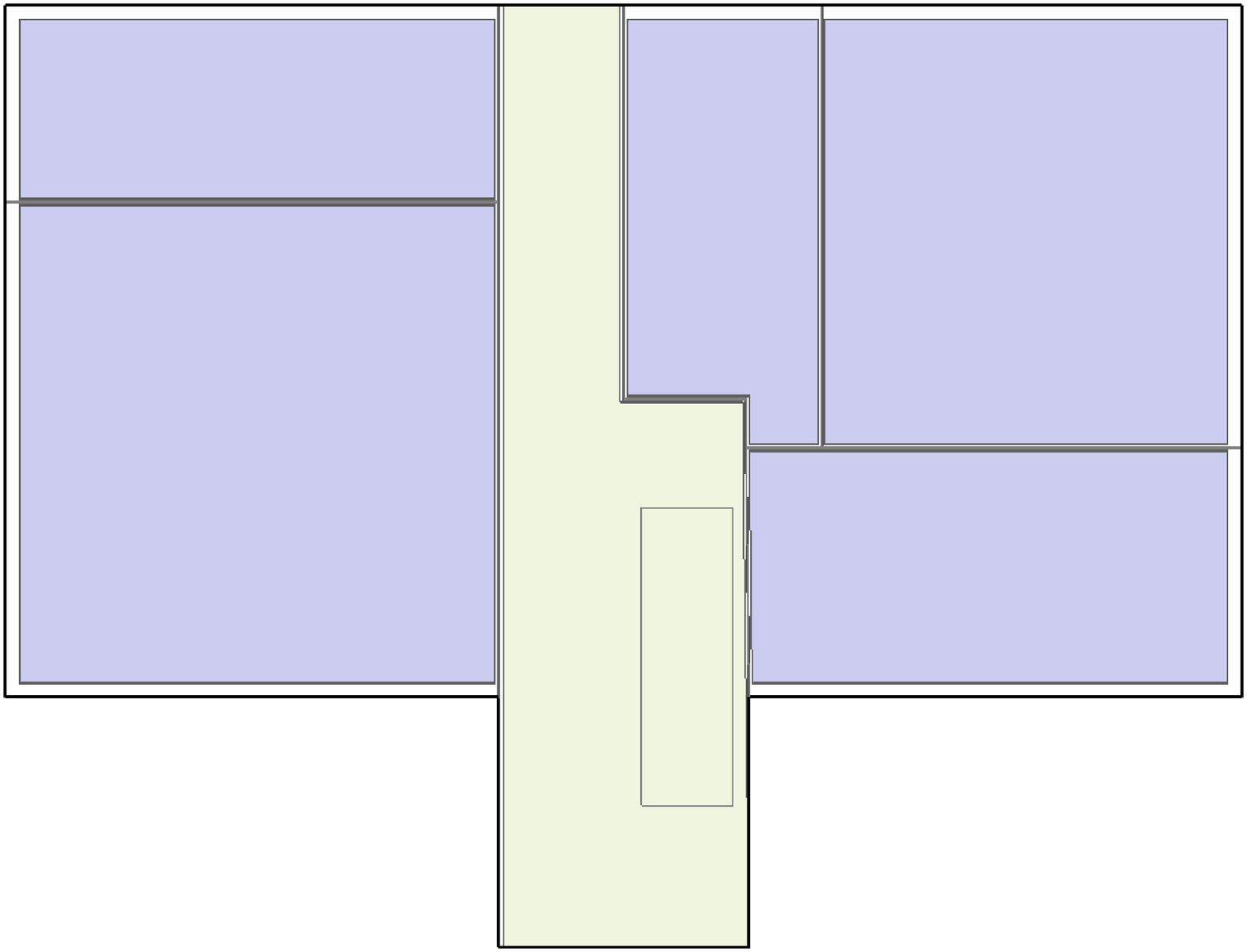}~~~~
	\includegraphics[width=0.3\columnwidth]{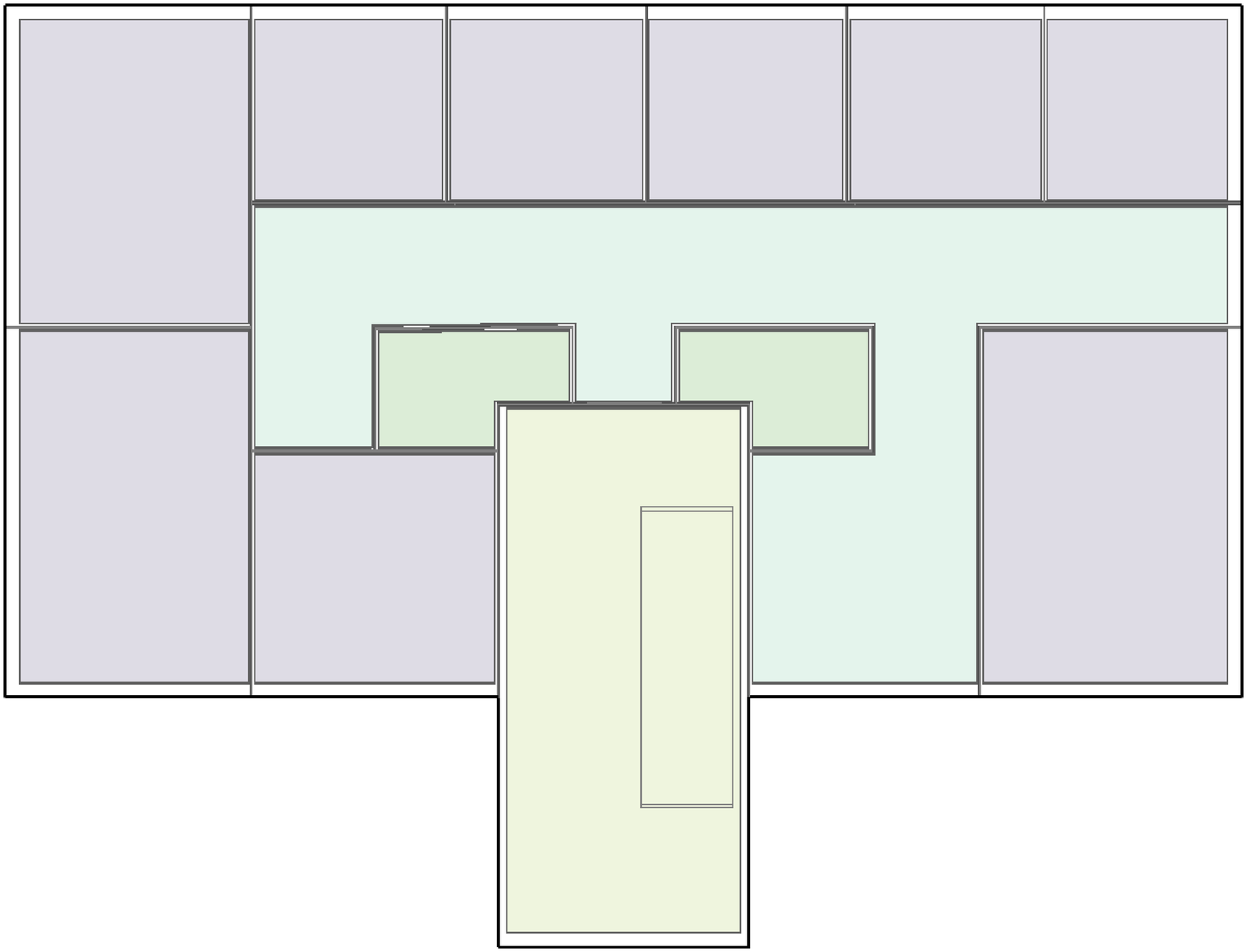}~~~~
	\includegraphics[width=0.3\columnwidth]{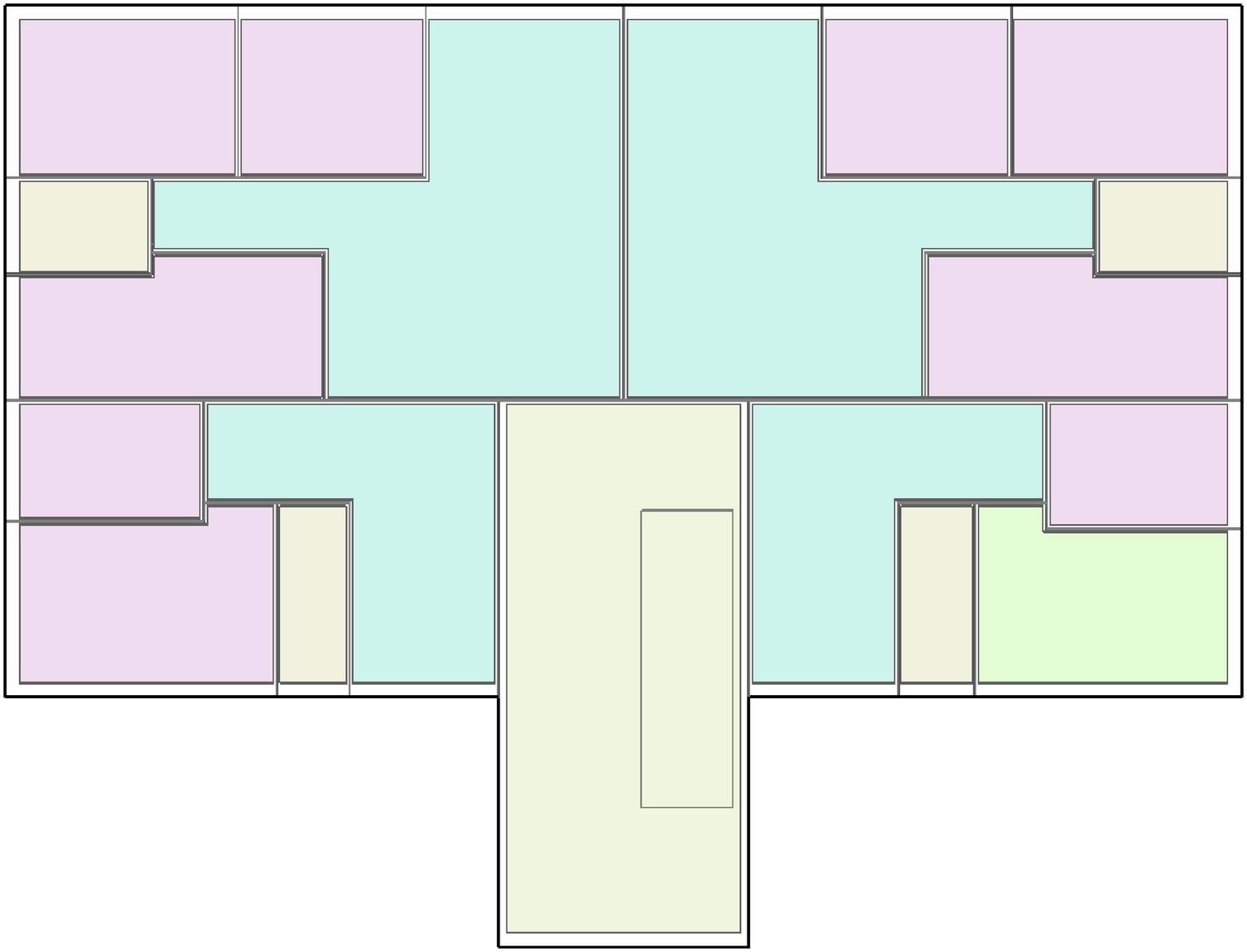}
	\caption{Plans of the building floors: commercial (left), office (middle), residential (right).}\label{fig:maps}
\end{figure}

The simulated test structure is an 8-floor building divided in 126 zones (Fig.~\ref{fig:building}) located in Turin, Italy. The first two floors are devoted to commercial and office activities, respectively, while the remaining floors are assigned to residential apartments. Plans of the three types of floors are reported in Fig.~\ref{fig:maps}.

\begin{figure}[H]
	\centering %
	\includegraphics[width=0.7\columnwidth]{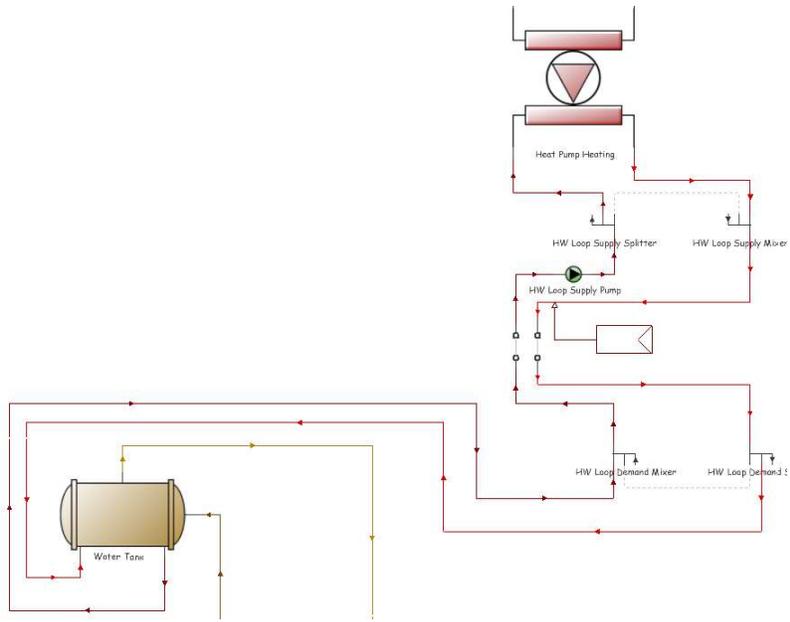}
	\caption{Scheme of the heating circuit.}\label{fig:heating_scheme}
\end{figure}

\begin{figure}[H]
	\centering %
	\includegraphics[width=0.4\columnwidth]{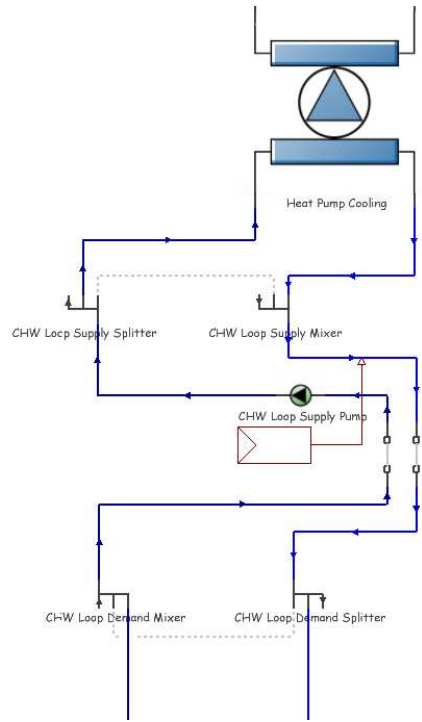}
	\caption{Scheme of the cooling circuit.}\label{fig:cooling_scheme}
\end{figure}

The building is equipped with a PV plant, an electrical storage system and two heat pumps used for heating and cooling purposes, respectively. The heating system is also equipped with a TES to enable hot water storage. Heating and cooling circuit schemes are shown in Fig.~\ref{fig:heating_scheme} and \ref{fig:cooling_scheme}. The building has been designed using DesignBuilder \cite{designbuilder} while simulations have been performed via EnergyPlus \cite{energyplus_13} connected to Matlab through the Building Controls Virtual Test Bed (BCVTB) \cite{BCVTB-Wetter11}. Building features are reported in Table~\ref{tab:building_characteristics}, while materials are summarized in Table~\ref{tab:building_materials}. The energy price time series $p(k)$ has been taken from the Italian electricity market \cite{GME}. The EnergyPlus simulation model is assumed as the real building. The sampling time $\tau_s$ is set to 10 minutes, while the horizon length used by the MPC is fixed to 12 hours, corresponding to 72 samples, i.e., $\lambda=72$ in Algorithm~\ref{algorithm}.

\begin{table}[htb]\centering
	\caption{Test building characteristics}\label{tab:building_characteristics}
	\begin{tabular}{llr} \hline
		\multicolumn{2}{l}{\textbf{Building Component}}                 & \textbf{Value} \\ \hline
		\multicolumn{2}{l}{Weather and Location}               & Turin (Italy)\\
		\multicolumn{2}{l}{Floor Area [$m^2$]}                 & 375 \\
		\multicolumn{2}{l}{Floor [\#]}                  & 8 \\
		\multicolumn{2}{l}{Zone [\#]}                   & 126 \\
		\multicolumn{2}{l}{Window to Wall Ratio [\%]}          & 20\\
		\multicolumn{2}{l}{Solar Transmittance [\%]}           & 30  \\
		\multirow{3}{*}{Internal Loads}  & Occupants [\#]      & 120 \\
		& Lighting [$W/m^2$]  & 4.00 \\
		& Equipment [$W/m^2$] & 3.25  \\ \hline
		\multirow{3}{*}{Heating: Heat Pump} & Heating Capacity [$kW$]& 389\\
		& Heating Power Consumption [$kW$]& 82\\
		& TES Volume [$m^3$]              & 8\\ \hline
		\multirow{2}{*}{Cooling: Heat Pump}     & Cooling Capacity [$kW$]& 389\\
		& Cooling Power Consumption [$kW$]& 82\\ \hline
		\multirow{3}{*}{Electric Energy Storage} & Capacity [$kWh$]& 28\\
		& Max Charging Rate [$kW$]& 24\\
		& Max Discharging Rate [$kW$] & 24\\ \hline
		\multirow{3}{*}{PV Plant} & Panels [\#] & 5 \\
		& Plant Surface Area [$m^2$]& 100\\
		& Plant Peak Power [$kW$]& 10\\ \hline

		\hline
	\end{tabular}
\end{table}
\begin{table}[H]\centering
	\caption{Test building construction materials (name/thickness [mm])}\label{tab:building_materials}
	\begin{small}
		\begin{tabular}{l|r|r|r|r|r|}
			\cline{2-3}
			& \textbf{External Walls~~~~~~}& \textbf{Internal Walls~~}\\ \cline{1-3}
			\multicolumn{1}{|c|}{\textbf{Outside Layer}}& Brickwork/100           & Gypsum plaster/13   \\ \cline{1-3} \multicolumn{1}{|c|}{\textbf{Layer 2}}      &
			Extruded polystyrene/80 & Brickwork/10            \\ \cline{1-3} \multicolumn{1}{|c|}{\textbf{Layer 3}}      & Concrete block/100      & Gypsum plaster/13   \\
			\cline{1-3} \multicolumn{1}{|c|}{\textbf{Layer 4}}      & Gypsum plaster/15       &                   \\ \cline{1-3}
		\end{tabular}\medskip\\
		\begin{tabular}{l|r|r|r|r|r|}
			\cline{2-4}
			& \textbf{Floor~~~~~~ }   & \textbf{Roof~~~~~~ }  & \textbf{Windows~~~} \\ \cline{1-4}
			\multicolumn{1}{|c|}{\textbf{Outside Layer}}& Extruded polystyrene/30 & Plywood/10                   & Generic LoE/6      \\ \cline{1-4}
			\multicolumn{1}{|c|}{\textbf{Layer 2}}      & Cast concrete/300            & Glass wool/100              & Air/6               \\ \cline{1-4}
			\multicolumn{1}{|c|}{\textbf{Layer 3}}      &                   & Cast concrete/10            & Generic Clear/6      \\ \cline{1-4}
			\multicolumn{1}{|c|}{\textbf{Layer 4}}      &                   & Gypsum board/13 &                      \\ \cline{1-4}
		\end{tabular}
	\end{small}
\end{table}

\subsection{Winter season simulation - heating operation}\label{subsec:winter}

The thermal behavior of the building has been modeled as in Section~\ref{ss:mod} by means of the autoregressive model \eqref{eq:Tdyn}-\eqref{eq:Tdyn2}. To this purpose, it is appropriate to define the exogenous input vector $\be(k)$ as
$$
\be(k)=[T^A(k)~~ I(k)~~ I^2(k)~~ I(k) T^A(k)~~ \mathbf{G}'(k)]'
$$
where $T^A(k)$ is the outside air temperature, $I(k)$ is the solar irradiance and $\mathbf{G}(k)$ denotes the vector of internal heat gains of the building zones due to human occupancy, lighting and equipments.

\begin{figure}[H]
	\centering %
	\includegraphics[width=0.5\columnwidth]{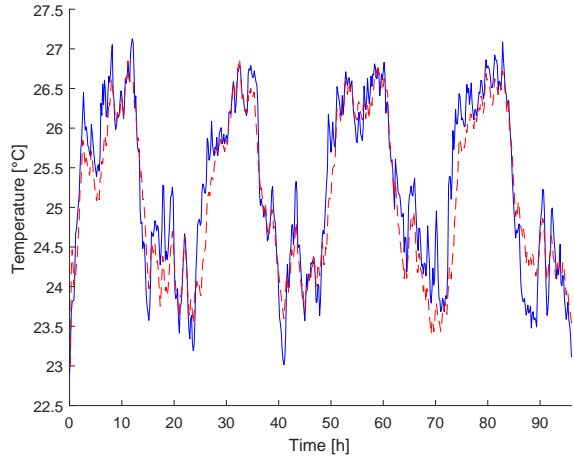}
	\caption{Thermal model identification. Comparison between real zone temperature (blue) and 6-hour ahead prediction (red) over the four-day validation period.}\label{fig:zone_heating_ident}
\end{figure}

For this model, an identification experiment has been performed over a time horizon of 18 days, which have been split in 14 days for estimation and 4 for validation. For each building zone, a coupled model involving all neighboring zones has been employed. The standard \emph{Best FIT index} (FIT) defined in \cite{Ljung,idtbx} has been used to assess the quality of the estimated model. The average value of the FIT index for all the identified building zones is 75\%, 68\% and 66\%, for 1, 6 and 12-hour ahead predictions, respectively. For a qualitative evaluation, in Fig.~\ref{fig:zone_heating_ident}, the 6-hour ahead prediction of the internal temperature of an office zone is compared with the real behavior during the model validation phase. 

Concerning the TES, model \eqref{eq:Hdynamics1}-\eqref{eq:Hdynamics2} has been identified, and the associated FIT index turns out to be over 90\% even for 12-hour ahead prediction. The PV plant has been identified by employing model \eqref{EQ:PVUSA2}.

Depending on the type of zone (commercial, office, residential), different comfort constraints are enforced. For commercial and office zones, temperature bounds are set to 20-24°C from 8:00 to 18:00, and 15-24°C elsewhen. Bounds for residential zones are 20-24°C from 7:00 to 9:00 and from 19:00 to 01:00, and 15-24°C otherwise. It is worthwhile to note that such bounds can be freely adjusted for each zone to adapt to real scenarios.

A 3-day simulation has been performed to evaluate the proposed control strategy. The price/volume DR requests in the considered period are assumed to be known one day in advance \cite{BCVE11} and are reported in Table~\ref{tab:DR_program_winter}. The identified model has been used by the MPC to generate fan input signals, HP setpoints and EES charging/discharging commands, while real measured data have been generated by the EnergyPlus simulation of the designed building.

\begin{table}[tb]\centering
	\caption{Winter season - DR program}\label{tab:DR_program_winter}
	\begin{tabular}{|c|c|c|c|c|}
		\hline ~ &$h_j$& $\mu_j$&  $S_j$ [$kWh$]&  $R_j$ [\euro]\\\hline %
		$\mathcal{R}_1$ & 107 & 5 & 37 & 3.6\\\hline %
		$\mathcal{R}_2$ & 178 & 6 & 40 & 4.4\\\hline %
		$\mathcal{R}3$ & 325 & 3 & 6 & 2.8\\\hline %
	\end{tabular}
\end{table}

At this stage, perfect knowledge of exogenous inputs is assumed. Such an assumption, although not realistic, has been enforced to better assess the quality of the proposed technique. Simulations in presence of uncertain forecasts will be analyzed in detail in Section~\ref{subsec:uncertainty}.

\begin{figure}[H]
	\centering %
	\includegraphics[width=0.5\columnwidth]{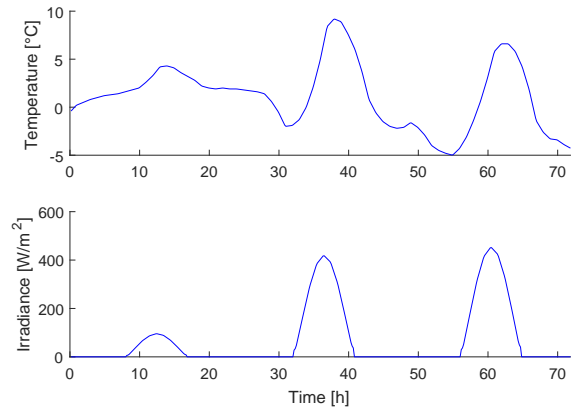}
	\caption{Heating operation mode simulation. Meteorological inputs. Top: Outdoor temperature. Bottom: Solar irradiance.}\label{fig:T_I_heating}
\end{figure}

\begin{figure}[H]
	\centering %
	\includegraphics[width=0.6\columnwidth]{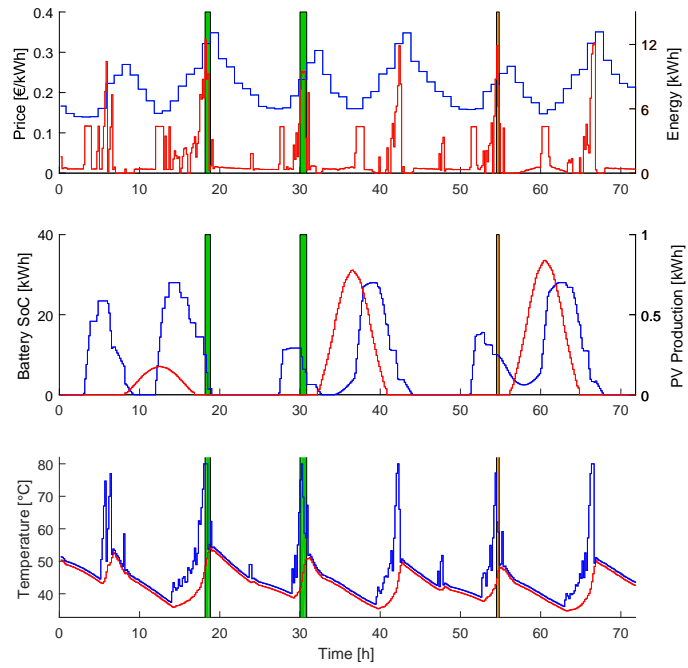}
	\caption{Heating operation mode simulation. DR requests are depicted in green if fulfilled, in orange otherwise. Top: energy price (blue) and building energy consumption per time step (red). Middle: EES state of charge (blue) and energy provided by the PV plant (red). Bottom: TES internal temperature (red) and HP setpoint (blue).}\label{fig:consumption_AD_heating}
\end{figure}

\begin{figure}[H]
	\centering %
	\includegraphics[width=0.6\columnwidth]{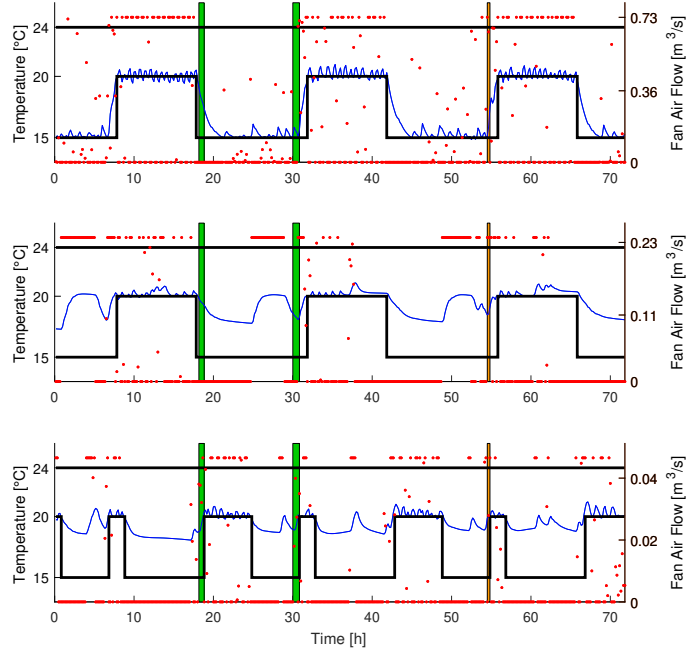}
	\caption{Heating operation mode simulation. DR requests are depicted in green if fulfilled, in orange otherwise. Zone internal temperature (blue), comfort bounds
		(black), fan speed (red dots). Top: commercial zone. Middle: office zone. Bottom: residential zone. }\label{fig:zone_heating_MPC}
\end{figure}

In Fig.~\ref{fig:T_I_heating}, outdoor temperature $T^A$ and solar irradiance $I$ for the considered days are reported, while in Fig.~\ref{fig:consumption_AD_heating}, TES temperature $T^{TES}$, HP setpoint $T^{HP_H}_0$, EES state of charge $E^{EES}$, energy drawn from the PV plant $W^{PV}$, unit energy price $p$ and total energy drawn from the grid $W$ are shown. In Fig.~\ref{fig:zone_heating_MPC}, the temperature profiles $T_i(k)$ of three different sample zones are depicted along with the required comfort bounds $[\underline T_i,\overline T_i]$ and fan actuation $v_i$. It can be noted that the zone temperatures lie within the comfort bounds with small deviations. Indeed, the average bound violation of the worst performing zone over three days is about 0.13°C, see Table~\ref{tab:winter_results}. Notice that, since exact input forecasts are assumed, such infringements are due to modeling errors. This demonstrates the ability of the bound relaxation technique in Section~\ref{subsec:manage_constr} to manage unpredictable bound violations by pushing the zone temperatures back within the prescribed bounds in a short time after the violations occur. It is worthwhile to recall that keeping air temperature within hard comfort bounds at all times is unfeasible for any conceivable control strategy due to modeling errors and/or unpredictable disturbances. In Table~\ref{tab:winter_results}, numerical values of total cost, DR rewards, and comfort bound violations are reported.

\begin{table}[H]\centering
	\caption{Heating operation mode - Simulation results over 3 days}\label{tab:winter_results}
	\begin{tabular}{|l|c|}\hline
		Cost without DR  [\euro]     &  136.14      \\\hline %
		No. of fulfilled DR requests &  2     \\\hline %
		DR reward [\euro]            &  8.00     \\\hline %
		Overall cost with DR [\euro] &  128.14     \\\hline %
		Worst zone average bound violation [°C] &  0.139     \\\hline %
	\end{tabular}
\end{table}

To further assess the performance of the proposed procedure, a standard thermostatic controller with 0.5°C hysteresis has been used as a benchmark. Since thermostatic control is clearly unable to track step variations of the comfort constraints, preheating/precooling is performed with suitable advance in order to provide a fair comparison. Moreover, when thermostatic control is in place, the EES is operated in the following fashion: if the power currently produced by the PV panels exceeds that needed by the HP, the surplus is stored in the battery, otherwise the controller draws power from the battery in the first place, and then from the grid. Since DR requests cannot be handled by thermostatic control, simulations in absence of DR programs have been performed for the sake of fairness. 

In Table~\ref{tab:winter_comparison}, the total cost is summarized for both strategies along with the average bound violation of the worst performing zone. In the considered days, the total cost achieved by adopting the proposed control technique turns out to be 25.49\% less than that obtained by the benchmark.

\begin{table}[H]\centering
	\caption{Heating operation mode simulation - Comparison with thermostatic control (without DR)}\label{tab:winter_comparison}
	\begin{tabular}{|l|c|c|}
		\hline ~ & Proposed MPC & Thermostatic Control\\\hline %
		Overall cost         [\euro] & 136.64  &  183.39  \\\hline %
		Worst zone average bound violation [°C] &  0.139 &  0.142  \\\hline %
	\end{tabular}
\end{table}

\subsection{Summer season simulation - cooling operation}\label{subsec:summer}

As described in Section~\ref{subsec:cooling_mode}, the problem structure for cooling mode is similar to the heating case, with the exception that no TES is assumed to be present. The identified models used under this condition have been obtained in a similar manner. A model for the dynamics of the fluid temperature at the HP inlet ($T^{HP_{C}}_{in}$) is identified according to \eqref{eq:Cdynamics1}-\eqref{eq:Cdynamics2}, and the resulting FIT index turns out to be over 80\% for 12-hour ahead predictions. This step is not explicitly needed for the heating case since, according to \eqref{eq:HPH}, the fluid temperature at HP inlet is the same as the TES temperature $T^{TES}$.

For commercial and office zones, comfort bounds are set to 22-24°C from 8:00 to 18:00, and 22-28°C elsewhen. Bounds for residential zones are 22-24°C from 7:00 to 9:00 and from 19:00 to 01:00, and 22-28°C otherwise.

\begin{figure}[H]
	\centering %
	\includegraphics[width=0.6\columnwidth]{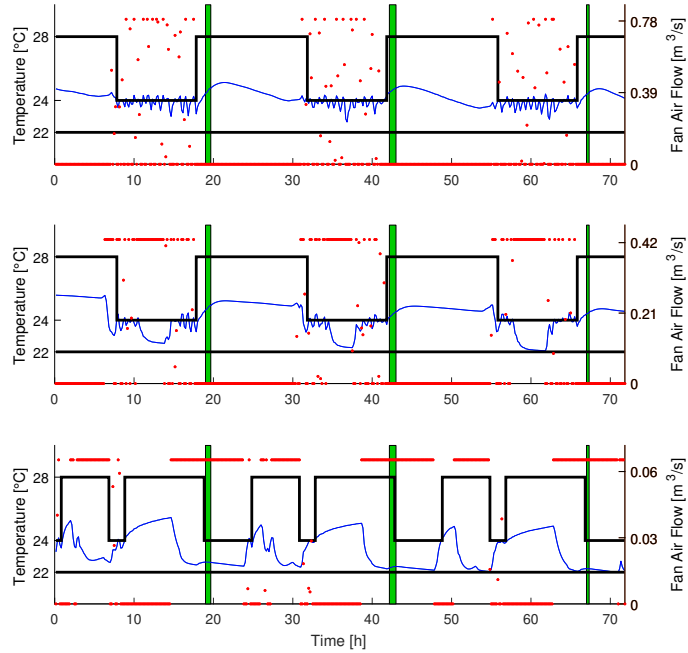}
	\caption{Cooling operation mode simulation. Zone internal temperature (blue), comfort bounds (black), fan speed (red dots) and fulfilled DR requests (green). Top:
		commercial zone. Middle: office zone. Bottom: residential zone. }\label{fig:zone_cooling_MPC}
\end{figure}

\begin{figure}[H]
	\centering %
	\includegraphics[width=0.6\columnwidth]{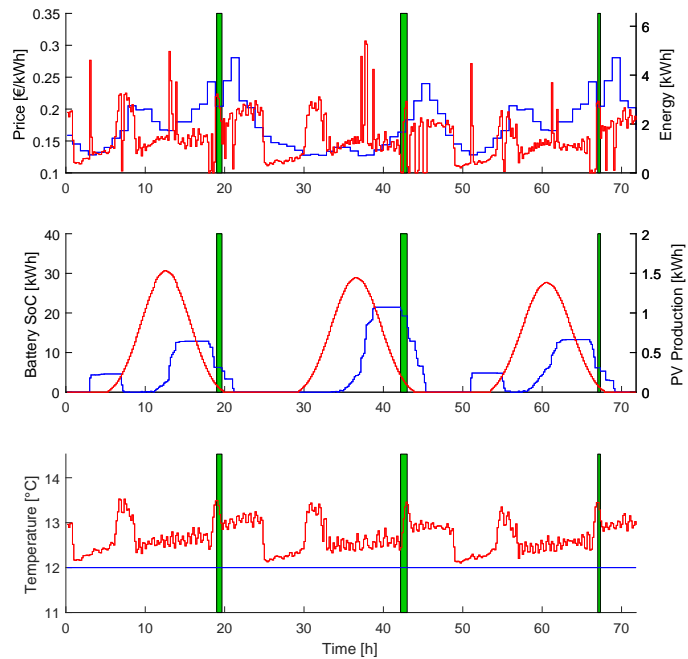}
	\caption{Cooling operation mode simulation. DR requests are depicted in green if fulfilled, in orange otherwise. Top: energy price (blue) and building energy consumption per time step (red). Middle: EES state of charge (blue) and energy provided by the PV plant (red). Bottom: HP setpoint (blue) and fluid temperature at HP inlet (red).}\label{fig:consumption_AD_cooling}
\end{figure}

In Table~\ref{tab:summer_results} and Fig.~\ref{fig:zone_cooling_MPC}-\ref{fig:consumption_AD_cooling}, the results of a three-day simulation are summarized. In Table~\ref{tab:summer_comparison}, the comparison with the thermostatic controller is shown. As for the winter season, the proposed MPC reduces the cost by 35.67\% with respect to the benchmark, still maintaining similar zone comfort.

\begin{table}[H]\centering
	\caption{Cooling operation mode - Simulation results over 3 days}\label{tab:summer_results}
	\begin{tabular}{|l|c|}\hline
		Cost without DR  [\euro]     &  79.68     \\\hline %
		No. of fulfilled DR requests &  3         \\\hline %
		DR reward [\euro]            &  6.40      \\\hline %
		Overall cost with DR [\euro] &  73.28     \\\hline %
		Worst zone average bound violation [°C] &  0.145    \\\hline %
	\end{tabular}
\end{table}

\begin{table}[H]\centering
	\caption{Cooling operation mode simulation - Comparison with thermostatic control (without DR)}\label{tab:summer_comparison}
	\begin{tabular}{|l|c|c|}
		\hline ~ & Proposed MPC & Thermostatic Control\\\hline %
		Overall cost         [\euro] & 76.74  & 119.29   \\\hline %
		Worst zone average bound violation [°C] & 0.156 & 0.169  \\\hline %
	\end{tabular}
\end{table}

\subsection{Performance analysis under uncertainties}\label{subsec:uncertainty}

Up to this point, exact forecasts of the exogenous inputs $\be$ have been considered. Such an assumption is not acceptable in a realistic scenario for weather variables and zone internal gains. On the contrary, it is reasonable to assume that energy price is exactly known one day in advance, as well as the daily DR program \cite{BCVE11}.
For these reasons, simulations have been performed to evaluate how the proposed technique is sensitive to forecasting errors in outdoor temperature, solar irradiance and internal gains.

In a real scenario, predictions of external temperature are provided by national weather forecast services. Outdoor temperature forecasting errors have been simulated as follows. First, a second-order autoregressive model has been used to compute the noise signal
\begin{equation}\label{eq:AR_unc}
d(k)=\alpha_1 \,d(k-1) + \alpha_2 \,d(k-2) + \eps(k),\quad k=0,\ldots
\end{equation}
where $\alpha_1$ and $\alpha_2$ are the model coefficients and $\eps$ is a zero-mean Gaussian distributed random variable. Notice that it is quite common to use autoregressive systems to model temperature uncertainty, see e.g., \cite{Darivianakis17}. Then, denoting by $T_{true}^A$ the real temperature profile, the temperature forecast $\widehat T^A$ to be used by the MPC has been obtained as
\begin{equation}\label{eq:imbutizzazione}
\widehat T^A(l)= T_{true}^A(l)+d(l)\,\frac{l-k}{\lambda},\quad l=k,\ldots,k+\lambda\ .
\end{equation}
Notice that $\widehat T^A(k)=T_{true}^A(k)$, since the outdoor temperature is measured. On the other hand, the term $\frac{l-k}{\lambda}$ takes into account the fact that forecasting accuracy decreases with time. State-of-the-art models for temperature forecasting are able to generate day-ahead estimates with an error of about 2°C \cite{Feldmann15}. Hence,  the coefficients in \eqref{eq:AR_unc} have been chosen conservatively in order obtain a maximum error of 3°C over a 12-hour ahead prediction.

\begin{figure}[H]
	\centering %
	\includegraphics[width=0.65\columnwidth]{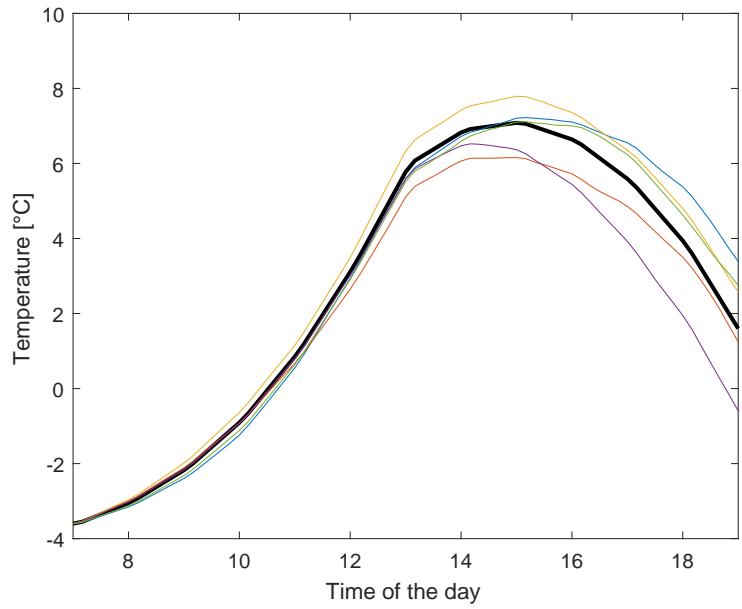}
	\caption{Real outdoor temperature (black thick line) and five simulated forecasts (colored thin lines) at a given time.}\label{fig:ext_temp_unc}
\end{figure}

To test the behavior of the proposed control law, a number of uncertain input profiles have been generated and used in simulations. In Fig.~\ref{fig:ext_temp_unc}, five different 12-hour ahead forecasts are reported, along with the true temperature profile. 

\begin{figure}[H]
	\centering %
	\includegraphics[width=0.65\columnwidth]{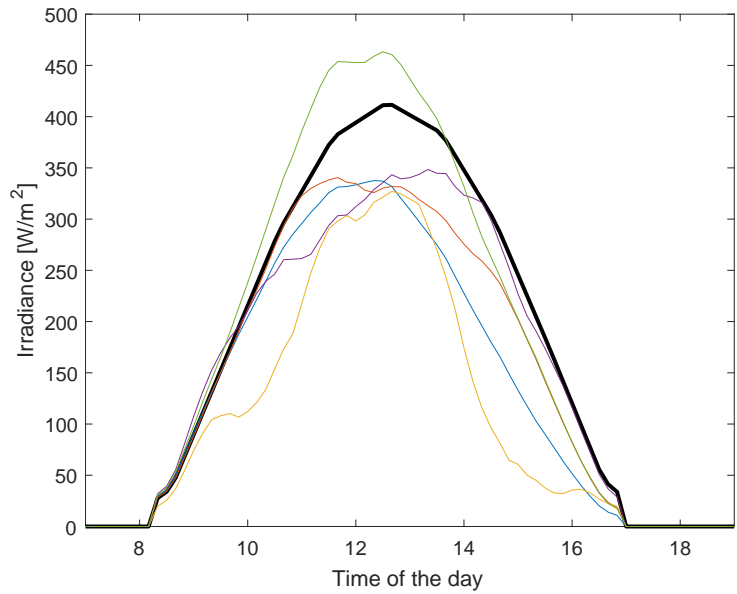}
	\caption{Real solar irradiance (black thick line) and five simulated forecasts (colored thin lines) at a given time.}\label{fig:irradiance_unc}
\end{figure}

Predictions of solar irradiance have been obtained by multiplying the real irradiance by a signal generated as in \eqref{eq:AR_unc}. In Fig.~\ref{fig:irradiance_unc}, different profiles of solar irradiance predictions are reported along with the actual signal.

\begin{figure}[H]
	\centering %
	\includegraphics[width=0.65\columnwidth]{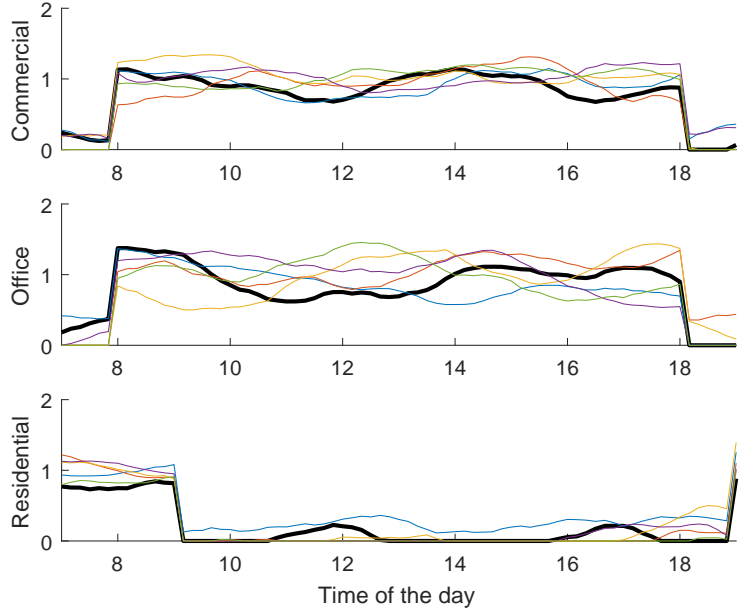}
	\caption{Real internal gains (black thick line) and five simulated forecasts (colored thin lines) at a given time. Top: Commercial zone. Middle: Office zone. Bottom: Residential zone.}\label{fig:ig_unc}
\end{figure}

To obtain forecasts on zone internal gains, a noise signal as in \eqref{eq:AR_unc} has been added to the real internal gain, by saturating the perturbed signal to 0. In Fig.~\ref{fig:ig_unc}, the true internal gain is depicted for each zone type along with five forecasts. 

\begin{figure}[H]
	\centering %
	\includegraphics[width=0.65\columnwidth]{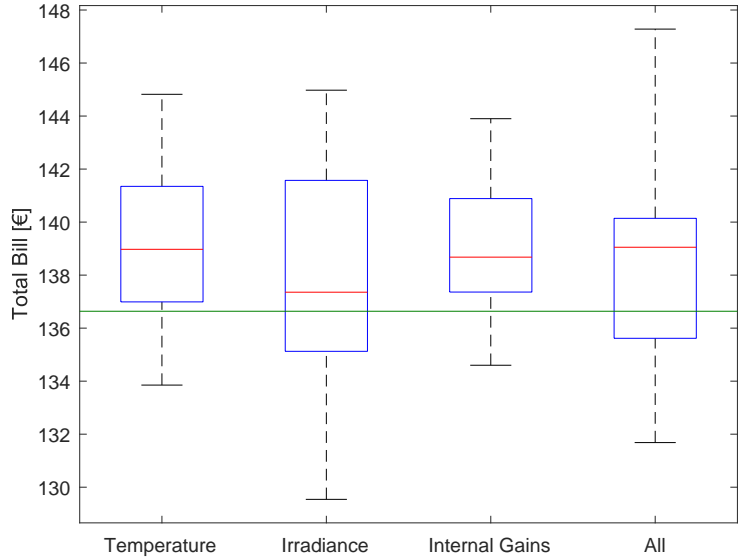}
	\caption{Box plot of the total cost in presence of uncertain forecasts over 20 simulations. The three sources of uncertainty are considered once at a time and at the same time. The green line refers to the result for exact forecasts.}\label{fig:uncertain_box}
\end{figure}

Twenty simulations in heating mode have been generated for each uncertain input. In
Table~\ref{tab:uncertain_result}, the worst-case value of the total cost and of the average comfort bound violation of the worst performing zone are reported. Inaccurate forecasts are considered both individually and combined. The box plots related to the overall cost for the considered scenarios are shown in Fig.~\ref{fig:uncertain_box}.

It can be noticed that the presence of inaccurate forecasts may raise the overall cost up to 7.79\% with respect to the nominal case, while the difference on comfort is negligible. The latter fact shows the ability of the control algorithm to minimize the thermal discomfort when unexpected situations occur.

\begin{table}[H]\centering%
	\caption{Simulation results for exact and inaccurate forecasts (worst-case over 20 realizations)}\label{tab:uncertain_result}
	\begin{tabular}{|l|c|c|c|}
		\hline ~ & Overall cost [\euro] & \parbox[t]{3.5cm}{\centering Worst zone average\\[-2mm] bound violation [°C]\\[-4mm]~}\\\hline %
		Exact forecast & 136.64 & 0.139\\\hline %
		\parbox[t]{4.1cm}{Uncertain forecast on \\[-2mm] external temperature \\[-4mm]} & 144.82 & 0.141\\\hline %
		\parbox[t]{3.7cm}{Uncertain forecast on \\[-2mm] solar irradiance \\[-4mm]} & 144.98 & 0.144\\\hline %
		\parbox[t]{3.7cm}{Uncertain forecast on \\[-2mm] zone internal gains \\[-4mm]}  & 143.90 & 0.144\\\hline %
		\parbox[t]{3.7cm}{Uncertain forecast on\\[-2mm]all three inputs\\[-4mm]}  &  147.28 &  0.146\\\hline %
	\end{tabular}
\end{table}

\section{Discussion}\label{sec:discussion}
In this section, the results of the simulation experiments are analyzed and discussed.

Let us first consider the simulation reported in Section~\ref{subsec:winter}, corresponding to winter season and perfect knowledge of exogenous inputs. The identified models are able to predict the system dynamics accurately. As expected, the performance indexes decrease as the prediction horizon grows, yet they remain on acceptable values. Moreover, it is worth remembering that in a receding horizon approach, predictions are updated at each time step, thus making it possible for the control algorithm to adapt against inaccuracies on long-term forecasts.

Concerning the control system simulation in Fig.~\ref{fig:consumption_AD_heating}, it can be observed that the EES and the TES are charged and discharged twice a day in order to take advantage of electricity price fluctuations. Moreover, it is apparent that both the EES and the TES play an important role in order to satisfy DR requests. From Fig.~\ref{fig:consumption_AD_heating} and \ref{fig:zone_heating_MPC}, it is apparent that the last DR request is not fulfilled, meaning that consumption reduction during the DR interval is not deemed profitable by the controller. In Fig.~\ref{fig:zone_heating_MPC}, one can see that the comfort bounds are mostly satisfied. In this case, since exact forecasts are assumed, bound violations are only due to discrepancies between the real system and the identified model. However, the magnitude of such bound violations is almost negligible. In fact, as reported in Table~\ref{tab:winter_results}, the average bound violation for the worst zone is less than 0.14°C. The proposed control technique leads to a total cost which is 25.49\% less than that obtained by standard thermostatic rules, as reported in Table~\ref{tab:winter_comparison}. Since the benchmark cannot exploit the knowledge of DR requests, no DR program has been assumed in this case to ensure a fair comparison. Of course, the gap between the two strategies in the presence of DR programs is larger.

Since the assumption of exact forecasts is unrealistic, the cost associated to the benchmark controller is also compared with that obtained by the MPC under inaccurate input predictions. To this purpose, Table~\ref{tab:uncertain_result} and Fig.~\ref{fig:uncertain_box} show that the MPC cost under uncertain forecasts does not change significantly w.r.t. the nominal one. In fact, the maximum cost under inaccurate forecasts is less than 8\% greater than that obtained for exact forecasts. This is a consequence of the receding horizon strategy, which implies that only the prediction errors occurring in the near future significantly affect the performance. This makes the proposed approach intrinsically robust to uncertainty, since short-term forecasts are usually quite accurate.

In addition to affecting the total cost, uncertainty also plays a role in the preservation of zone temperature comfort bounds. However, this effect is mitigated by the MPC implementation described in Section~\ref{subsec:manage_constr}, as it is clear from Table~\ref{tab:uncertain_result}, which shows that only a slight increment of discomfort occurs in the presence of unreliable forecasts. 

Similar considerations can be made for the results obtained in cooling operation, i.e., in summer season.

A final remark concerns the computational burden associated with the controller implementation. For the considered 126-zone building, the time required for computations at each step (i.e., every 10 minutes) is about 20 seconds on a standard PC\footnote{Computations have been performed using CPLEX \cite{ilog} to solve the LPs, on an Intel Core i7-3770 at 3.40 GHz with 8 GB of RAM.}, showing that the proposed algorithm can be efficiently adopted in large-scale applications.

\section{Conclusion and future research}\label{sec:conclusions}
In this paper, the problem of optimizing the operation cost (i.e., the electrical energy bill) of a building integrating a centralized HVAC, thermal and electrical storage facilities, and PV generation has been addressed. Participation in a DR program has also been considered.  The proposed approach exploits a receding horizon control MPC strategy involving at each step the solution of an LP and of a MILP with a number of integer variables equal to the number of DR requests in the prediction horizon. The procedure can be fruitfully applied to large size buildings. Experimental validation using a realistic simulation framework has been carried out.

Further studies will address more complex setups in which the building is considered as a microgrid in its own right, including electric vehicles, appliances and other kinds of loads. Special attention to active/reactive power flow should be paid in this case in order to guarantee the satisfaction of electrical constraints. More general DR scenarios will also be addressed, such as the presence of incentives for keeping energy consumption above a minimum, which can be efficiently achieved using storage facilities.
Finally, different methods for handling uncertainty, like chance constraints or scenario-based approaches, will be investigated and compared with the method proposed in this paper. To this purpose, suitable problem formulations and related relaxations will be studied to enable such techniques to manage large-scale buildings.

\section*{References}

\bibliographystyle{elsarticle-num}
\bibliography{./buildings}

\end{document}